\documentclass[final,leqno]{siamltex}
\usepackage{amsmath}
\usepackage{amssymb}

\input psfig.sty

\newcommand{\p}{\partial}
\newcommand{\og}{\omega}

\newcommand{\fl}[2]{\frac{#1}{#2}}

\newcommand{\nn}{\nonumber}
\newcommand{\ap}{\alpha}

\newcommand{\ld}{\lambda}

\newcommand{\gm}{\gamma}

\newcommand{\tht}{\theta}
\newcommand{\ift}{\infty}
\newcommand{\vep}{\varepsilon}

\newcommand{\sg}{\sigma}

\newcommand{\btu}{\Delta}

\newcommand{\be}{\begin{equation}}
\newcommand{\ee}{\end{equation}}
\newcommand{\ba}{\begin{array}}
\newcommand{\ea}{\end{array}}
\newcommand{\bea}{\begin{eqnarray}}
\newcommand{\eea}{\end{eqnarray}}
\newcommand{\beas}{\begin{eqnarray*}}
\newcommand{\eeas}{\end{eqnarray*}}

\newtheorem{remark}{Remark}[section]

 \newcommand{\bx}{{\bf x} }

\title{Computing Ground States of Spin-1 Bose-Einstein Condensates
 by the Normalized Gradient Flow}
\author{Weizhu Bao
\thanks{Department of Mathematics and Center for Computational Science
and Engineering, National University of Singapore, Singapore
117543 ({\it bao@math.nus.edu.sg}, URL:
http://www.math.nus.edu.sg/\~{}bao/)}  \and Fong Yin Lim
\thanks{Department of Mathematics and Center for Computational Science
and Engineering, National University of Singapore, Singapore
117543 ({\it fongyin.lim@nus.edu.sg})}. }

\date{}
\begin{document}

\maketitle


\begin{abstract}
In this paper, we propose an efficient and accurate numerical
method for computing the ground state of spin-1 Bose-Einstein
condensates (BEC) by using the normalized gradient flow or
imaginary time method. The key idea is to find a third projection
or normalization condition based on the relation between the chemical
potentials so that the three projection parameters used in the
projection step of the normalized gradient flow are uniquely
determined by this condition as well as the other two physical
conditions given by the conservation of total mass and total
magnetization. This
allows us to successfully extend the most popular and powerful
normalized gradient flow or imaginary time method for computing
the ground state of single component BEC to compute the ground
state of spin-1 BEC. An efficient and accurate discretization scheme, the
backward-forward Euler sine-pseudospectral method (BFSP), is
proposed to discretize the normalized gradient flow. Extensive
numerical results on ground states of spin-1 BEC with
ferromagnetic/antiferromagnetic interaction and harmonic/optical
lattice potential in one/three dimensions are reported to
demonstrate the efficiency of our new numerical method.

\end{abstract}



\pagestyle{myheadings} \markboth{W. Bao and F. Y. Lim} {Computing
ground states of spin-1 BEC by normalized gradient flow}

\section{Introduction}
\label{s1}
\setcounter{equation}{0}

Research in low temperature dilute atomic quantum gases remains
active for more than ten years
after the experimental realizations of Bose-Einstein condensation
(BEC) in alkali atomic gases
in 1995 \cite{Wieman_Cornell, Hulet, Ketterle}. Extensive theoretical
and experimental studies
have been carried out to investigate various novel phenomena of the
condensates.
In earlier BEC experiments, the atoms were confined in magnetic trap
\cite{Wieman_Cornell, Hulet, Ketterle},
in which the spin degrees of freedom is frozen. The particles are
described by a scalar model and the wavefunction of the particles is
governed by the Gross-Pitaevskii equation (GPE) within the mean-field
 approximation
\cite{Dalfovo, Gross, Pitaevskii}. In recent years, experimental
achievement of spin-1 and spin-2
condensates \cite{Barrett, Ketterle2, Ketterle3, Ketterle4, Ketterle6}
offers new
regimes to study various quantum phenomena that are generally absent in
a single component condensate.
The spinor condensate is achieved experimentally when an optical trap,
instead of a magnetic trap,
is used to provide equal confinement for all hyperfine states.

The theoretical studies of spinor condensate have been carried out
in several papers since the achievement of it in experiments
\cite{Ho, Law, Ohmi, Ketterle5}. In contrast to single component
condensate, a spin-$F$ ($F\in {\Bbb N}$) condensate is described
by a generalized coupled GPEs which consists of $2F+1$ equations,
each governing one of the $2F+1$ hyperfine states $(m_F = -F,
-F+1,..., F-1, F)$ within the mean-field approximation. For a
spin-1 condensate, at temperature much lower than the critical
temperature $T_c$, the three-components wavefunction $\Psi(\bx,t)
= (\psi_1(\bx,t), \psi_0(\bx,t), \psi_{-1}(\bx,t)^T$ are well
described by the following coupled GPEs \cite{Ketterle5, You1,
You2, You3},
\begin{eqnarray}
\label{eq:CGPE_1}
i\hbar\,\p_t\psi_{1}(\bx,t)&=&\left[-\fl{\hbar^2}{2m}\nabla^2
+V(\bx)+
(c_0+c_2)\left(|\psi_1|^2+|\psi_0|^2\right)+(c_0-c_2)|\psi_{-1}|^2\right]
\psi_1\nn\\
&&+c_2\,\bar{\psi}_{-1}\,\psi_0^2, \\
 \label{eq:CGPE_2}
i\hbar\,\p_t\psi_{0}(\bx,t)&=&\left[-\fl{\hbar^2}{2m}\nabla^2
+V(\bx)+
(c_0+c_2)\left(|\psi_1|^2+|\psi_{-1}|^2\right)+c_0|\psi_{0}|^2\right]
\psi_0 \nn \\
&&+2c_2\,\psi_{-1}\,\bar{\psi}_{0}\,\psi_1, \\
\label{eq:CGPE_3}
i\hbar\,\p_t\psi_{-1}(\bx,t)&=&\left[-\fl{\hbar^2}{2m}\nabla^2
+V(\bx)+
(c_0+c_2)\left(|\psi_{-1}|^2+|\psi_0|^2\right)+(c_0-c_2)|\psi_{1}|^2\right]
\psi_{-1} \nn\\
&&+c_2\,\psi_0^2\,\bar{\psi}_{1}.
\end{eqnarray}
Here, $\bx=(x,y,z)^T$ is the Cartesian coordinate vector, $t$ is
time, $\hbar$ is the Planck constant, $m$ is the atomic mass, and
$V(\bx)$ is the external trapping potential. When a harmonic trap
potential is considered, \be\label{poten}
V(\bx)=\frac{m}{2}(\og_x^2 x^2 +\og_y^2 y^2+ \og_z^2 z^2), \ee
 with $\og_x$, $\og_y$ and $\og_z$
being the trap frequencies in the $x$-, $y$- and $z$-direction,
respectively.
$\bar{f}$ and Re$(f)$ denote the conjugate and real part of the
function $f$, respectively. There are two atomic collision terms,
$c_0 = \frac{4\pi\hbar^2}{3m}(a_0+2a_2)$ and $c_2 =
\frac{4\pi\hbar^2}{3m}(a_2-a_0)$, expressed in terms of the
$s$-wave scattering lengths, $a_0$ and $a_2$, for scattering
channel of total hyperfine spin 0 (anti-parallel spin collision)
and spin 2 (parallel spin collision), respectively. The usual
mean-field interaction, $c_0$, is positive for repulsive
interaction and negative for attractive interaction. The
spin-exchange interaction, $c_2$, is positive for
antiferromagnetic interaction and negative for ferromagnetic
interaction. The wave function is normalized according to \be
\label{eq:Ns} \|\Psi\|^2:=\int_{{\Bbb R}^3} |\Psi(\bx,t)|^2\;d\bx=
\int_{{\Bbb R}^3} \sum_{l=-1}^1 |\psi_l(\bx,t)|^2\;d\bx
:=\sum_{l=-1}^1 \|\psi_l\|^2 =N, \ee where $N$ is the total number
of particles in the condensate.

By introducing the dimensionless variables: $t\to t/\og_m$ with
$\og_m=\min\{\og_x,\;\og_y,\;\og_z\}$, $\bx\to \bx \; a_s $ with
$a_s=\sqrt{\frac{\hbar}{m\og_m}}$, $\psi_l\to
\sqrt{N}\psi_l/a_s^{3/2}$ ($l=-1,0,1$), we  get the dimensionless
coupled GPEs from (\ref{eq:CGPE_1})-(\ref{eq:CGPE_3}) as
\cite{You2,ZhangY,Bao_Wang}:
\begin{eqnarray}
\label{GPEs80} i\p_t\psi_{1}(\bx,t)&=&\left[-\fl{1}{2}\nabla^2
+V(\bx)+ (\beta_n+\beta_s)\left(|\psi_1|^2+|\psi_0|^2\right)
+(\beta_n-\beta_s)|\psi_{-1}|^2\right]\psi_1\nn\\
&&+\beta_s\,\bar{\psi}_{-1}\,\psi_0^2,\\
\label{GPEs81} i\p_t\psi_{0}(\bx,t)&=&\left[-\fl{1}{2}\nabla^2
+V(\bx)+ (\beta_n+\beta_s)\left(|\psi_1|^2+|\psi_{-1}|^2\right)
+\beta_n|\psi_{0}|^2\right]\psi_0 \nn \\
&&+2\beta_s\,\psi_{-1}\,\bar{\psi}_{0}\,\psi_1,\\
\label{GPEs82} i\p_t\psi_{-1}(\bx,t)&=&\left[-\fl{1}{2}\nabla^2
+V(\bx)+ (\beta_n+\beta_s)\left(|\psi_{-1}|^2+|\psi_0|^2\right)
+(\beta_n-\beta_s)|\psi_{1}|^2\right]\psi_{-1} \nn\\
&&+\beta_s\,\psi_0^2\,\bar{\psi}_{1};
\end{eqnarray}
where $\beta_n=\frac{N\;c_0}{a_s^3\hbar\og_m}=\frac{4\pi
N(a_0+2a_2)}{3a_s}$,
$\beta_s=\frac{N\;c_2}{a_s^3\hbar\og_m}=\frac{4\pi
N(a_2-a_0)}{3a_s}$ and $V(\bx)=\frac{1}{2}(\gm_x^2 x^2$ $+\gm_y^2
y^2 +\gm_z^2 z^2)$ with $\gm_x=\frac{\og_x}{\og_m}$,
$\gm_y=\frac{\og_y}{\og_m}$ and $\gm_z=\frac{\og_z}{\og_m}$.
Similar as those in single component BEC
\cite{LS,BaoT,Bao3,Bao_Jaksch_Markowich}, in a disk-shaped
condensation, i.e. $\og_x\approx \og_y$ and $\og_z\gg\og_x$
($\Longleftrightarrow \gm_x=1$, $\gm_y\approx 1$ and $\gm_z\gg1$
with $\og_m=\og_x$), the 3D coupled GPEs
(\ref{GPEs80})-(\ref{GPEs82}) can be reduced to a 2D coupled GPEs;
and in a cigar-shaped condensation, i.e. $\og_y\gg \og_x$ and
$\og_z\gg\og_x$ ($\Longleftrightarrow \gm_x=1$, $\gm_y\gg 1$ and
$\gm_z\gg1$ with $\og_m=\og_x$), the 3D coupled GPEs
(\ref{GPEs80})-(\ref{GPEs82}) can be reduced to a 1D coupled GPEs.
Thus here we consider the dimensionless coupled GPEs in
$d$-dimensions ($d=1,2,3$):
\begin{eqnarray}
\label{eq:dCGPE_1} i\p_t\psi_{1}(\bx,t)&=&\left[-\fl{1}{2}\nabla^2
+V(\bx)+ (\beta_n+\beta_s)\left(|\psi_1|^2+|\psi_0|^2\right)
+(\beta_n-\beta_s)|\psi_{-1}|^2\right]\psi_1\nn\\
&&+\beta_s\,\bar{\psi}_{-1}\,\psi_0^2,\\
\label{eq:dCGPE_2} i\p_t\psi_{0}(\bx,t)&=&\left[-\fl{1}{2}\nabla^2
+V(\bx)+ (\beta_n+\beta_s)\left(|\psi_1|^2+|\psi_{-1}|^2\right)
+\beta_n|\psi_{0}|^2\right]\psi_0 \nn \\
&&+2\beta_s\,\psi_{-1}\,\bar{\psi}_{0}\,\psi_1,\\
\label{eq:dCGPE_3}
i\p_t\psi_{-1}(\bx,t)&=&\left[-\fl{1}{2}\nabla^2 +V(\bx)+
(\beta_n+\beta_s)\left(|\psi_{-1}|^2+|\psi_0|^2\right)
+(\beta_n-\beta_s)|\psi_{1}|^2\right]\psi_{-1} \nn\\
&&+\beta_s\,\psi_0^2\,\bar{\psi}_{1}.
\end{eqnarray}
In the equations above, $V(\bx)$ is a real-valued potential whose shape is
determined by the type of system under investigation,
$\beta_n\propto N$ and $\beta_s \propto N$ correspond to the
dimensionless mean-field (spin-independent) and spin-exchange
interaction, respectively. Three important invariants of
(\ref{eq:dCGPE_1})-(\ref{eq:dCGPE_3}) are the {\sl mass} (or
normalization) of the wave function  \be \label{norm1}
N(\Psi(\cdot,t)):=\|\Psi(\cdot,t)\|^2:= \int_{{\Bbb R}^d}
\sum_{l=-1}^1 |\psi_l(\bx,t)|^2\;d\bx\equiv N(\Psi(\cdot,0)) = 1,
\qquad t\ge0, \ee the {\sl magnetization} (with $-1\le M\le 1$)
\be \label{magn} M(\Psi(\cdot,t)):=\int_{{\Bbb R}^d}
\left[|\psi_1(\bx,t)|^2- |\psi_{-1}(\bx,t)|^2\right]d\bx\equiv
M(\Psi(\cdot,0))= M \ee and the energy per particle
\begin{eqnarray}
\label{energy} E(\Psi(\cdot,t))&=&\int_{{\Bbb
R}^d}\biggl\{\sum_{l=-1}^{1} \left(\frac{1}{2}|\nabla
\psi_l|^2+V(\bx)|\psi_l|^2\right)
+(\beta_n-\beta_s)|\psi_1|^2 |\psi_{-1}|^2 \nn\\
&&+\frac{\beta_n}{2}|\psi_0|^4
+\fl{\beta_n+\beta_s}{2}\Bigl[|\psi_1|^4+|\psi_{-1}|^4
+2|\psi_0|^2
\left(|\psi_1|^2+|\psi_{-1}|^2\right)\Bigr]\nn\\
&&+\beta_s\left(\bar{\psi}_{-1}\psi_0^2\bar{\psi}_{1}+
\psi_{-1}\bar{\psi}_0^2\psi_{1}\right)\biggr\}\; d\bx \equiv
E(\Psi(\cdot,0)), \qquad t\ge0.
\end{eqnarray}

A fundamental problem in studying BEC is to find the condensate
stationary states $\Phi(\bx)$, in particular the ground state
which is the lowest energy stationary state. In other words, the
ground state, $\Phi_g(\bx)$, is obtained from the minimization of
the energy functional subject to the conservation of total mass
and magnetization:
\begin{quote}
  Find $\left(\Phi_g \in S\right)$ such that
\end{quote}
  \begin{equation}\label{eq:minimize}
    E_g := E\left(\Phi_g\right) = \min_{\Phi \in S}
    E\left(\Phi\right),
  \end{equation}
where the nonconvex set $S$ is defined as \be \label{eq:S}
S=\left\{\Phi=(\phi_1,\phi_0,\phi_{-1})^T\ |\ \|\Phi\|=1, \
\int_{{\Bbb R}^d} \left[|\phi_1(\bx)|^2
-|\phi_{-1}(\bx)|^2\right]=M, \ E(\Phi)<\ift\right\}. \ee This is
a nonconvex minimization problem. When $\beta_n\ge0$ and
$\beta_n\ge|\beta_s|$ and $\lim_{|\bx|\to\ift} V(\bx)=\ift$, the
existence of a minimizer of the nonconvex minimization problem
(\ref{eq:minimize}) follows from the standard theory \cite{Simon}.
For understanding the uniqueness question note that $E(\ap\cdot
\Phi_g)=E(\Phi_g)$ for all $\ap=\left(e^{i\tht_1}, e^{i\tht_0},
e^{i\tht_{-1}}\right)^T$ with $\tht_1+\tht_{-1}=2\tht_0$. Thus
additional constraints have to be introduced to show the
uniqueness.

As derived in \cite{Bao_Wang}, by defining the Lagrangian \be
\label{lag} {\cal L}(\Phi,\mu,\ld):=E(\Phi) -\mu\left(\|\phi_1\|^2
+\|\phi_0\|^2 +\|\phi_{-1}\|^2 -1\right)-\ld
\left(\|\phi_1\|^2-\|\phi_{-1}\|^2-M\right), \ee we get the
Euler-Lagrange equations associated to the minimization problem
(\ref{eq:minimize}):
\begin{eqnarray}
\label{GPEs30} (\mu+\ld)\;\phi_{1}(\bx)&=&\left[-\fl{1}{2}\nabla^2
+V(\bx)+ (\beta_n+\beta_s)\left(|\phi_1|^2+|\phi_0|^2\right)
+(\beta_n-\beta_s)|\phi_{-1}|^2\right]\phi_1\nn\\
&&+\beta_s\,\bar{\phi}_{-1}\,\phi_0^2:=H_1\, \phi_1, \eea \bea
\label{GPEs31} \mu\;\phi_{0}(\bx)&=&\left[-\fl{1}{2}\nabla^2
+V(\bx)+ (\beta_n+\beta_s)\left(|\phi_1|^2+|\phi_{-1}|^2\right)
+\beta_n|\phi_{0}|^2\right]\phi_0 \nn \\
&&+2\beta_s\,\phi_{-1}\,\bar{\phi}_{0}\,\phi_1:=H_0\,\phi_0, \eea
\bea \label{GPEs32}
(\mu-\ld)\;\phi_{-1}(\bx)&=&\left[-\fl{1}{2}\nabla^2 +V(\bx)+
(\beta_n+\beta_s)\left(|\phi_{-1}|^2+|\phi_0|^2\right)
+(\beta_n-\beta_s)|\phi_{1}|^2\right]\phi_{-1} \nn\\
&&+\beta_s\,\phi_0^2\,\bar{\phi}_{1}:=H_{-1}\,\phi_{-1}.
\end{eqnarray}
Here $\mu$ and $\ld$ are the Lagrange multipliers (or chemical
potentials) of the coupled GPEs
(\ref{eq:dCGPE_1})-(\ref{eq:dCGPE_3}). In addition,
(\ref{GPEs30})-(\ref{GPEs32}) is also a nonlinear eigenvalue
problem with two constraints \bea \label{con11} &&\|\Phi\|^2:=
\int_{{\Bbb R}^d} |\Phi(\bx)|^2\;d\bx =\int_{{\Bbb R}^d}
\sum_{l=-1}^1 |\phi_l(\bx)|^2\;d\bx
:=\sum_{l=-1}^1 \|\phi_l\|^2=1,\\
\label{con22} &&\|\phi_{1}\|^2-\|\phi_{-1}\|^2:= \int_{{\Bbb R}^d}
\left[|\phi_1(\bx)|^2-|\phi_{-1}(\bx)|^2\right]d\bx =M. \eea In
fact, the nonlinear eigenvalue problem
(\ref{GPEs30})-(\ref{GPEs32}) can also be obtained from the
coupled GPEs (\ref{eq:dCGPE_1})-(\ref{eq:dCGPE_3}) by plugging
$\psi_l(\bx,t)=e^{-i\mu_l t}\phi_l(\bx)$ ($l=1,0,-1$) with
\be\label{chem8} \mu_1=\mu+\ld,\quad  \mu_0=\mu, \quad
\mu_{-1}=\mu-\ld \qquad \Longleftrightarrow \qquad
\mu_1+\mu_{-1}=2\mu_0. \ee
 Thus it is also called as
time-independent coupled GPEs. In physics literatures, any
eigenfucntion $\Phi$ of the nonlinear eigenvalue problem
(\ref{GPEs30})-(\ref{GPEs32}) under constraints (\ref{con11})
and (\ref{con22}), whose energy is larger than the energy of the
ground state is called as an excited state of the coupled GPEs
(\ref{eq:dCGPE_1})-(\ref{eq:dCGPE_3}).


A widely used numerical method for computing the ground state of a
single component condensate is the imaginary time method followed
by an appropriate discretization scheme \cite{Chiofalo,Bao_Du,Bao3}
to evolve the resulted gradient flow equation under normalization
of the wavefunction, which is mathematically justified by using
the normalized gradient flow \cite{Bao_Du,Bao3}. However, it is
not obvious that this most popular and powerful normalized
gradient flow (or imaginary time method) could be directly
extended to compute the ground state of spin-1 BEC. The reason is
that we only have two normalization conditions (i.e. the two
constraints: conservation of total mass and magnetization) which
are insufficient to determine the three projection constants for
the three components of the wavefunction used in the normalization
step. In physics literatures, the imaginary time method is still applied
to compute the ground state of spin-1 BEC through the introduction of
a random variable to choose the three projection parameters in the
projection step \cite{You2,ZhangY}. Of course, this is not a
determinate and efficient way to compute the ground state of
spin-1 BEC due to the choice of the random variable. Recently, Bao
and Wang \cite{Bao_Wang} have proposed a continuous normalized
gradient flow (CNGF) for computing the ground state of spin-1 BEC.
The CNGF is discretized by Crank-Nicolson finite difference method
with a proper and very special way to deal with the nonlinear
terms and thus the discretization scheme can be proved to be mass and
magnetization conservative and energy diminishing in the
discretized level \cite{Bao_Wang}.  However, at each time step, a
fully nonlinear system must be solved which is a little tedious
from computational point of view since  the CNGF is an
integral-differential equations (see details in
(\ref{cngf1})-(\ref{FPhi}))
which involves implicitly the
Lagrange multipliers in the normalized gradient flow evolution
\cite{Bao_Wang}. The aim of this paper is to introduce a third
normalization condition based on the relation between the chemical
potentials of spin-1 BEC, in addition to the two existing
normalization conditions given by the conservation of the total mass
and magnetization. Thus we can completely determine the three projection
constants used in the normalization step for the normalized gradient flow.
This allows us to develop the most popular and powerful
normalized gradient flow
or imaginary time method to compute the ground state of spin-1 BEC.

The paper is organized as follows. In section \ref{s2},  the
normalized gradient flow is constructed by introducing the third
projection or normalization condition for computing the ground
state of spin-1 BEC.
 In section \ref{s3}, the backward-forward
Euler sine-pseudospectral method (BESP) is presented to discretize
the normalized gradient flow. In section \ref{s4}, ground states
of spin-1 BEC are reported with ferromagnetic/antiferromagnetic
interaction and  harmonic/optical lattice potential in one/three
dimensions, respectively.  Finally, some conclusions are drawn in
section \ref{s5}.

\section{The normalized gradient flow}
\label{s2}
\setcounter{equation}{0}

In this section,  we will construct the normalized gradient flow
for computing the ground state of spin-1 BEC by introducing the
third normalization condition.

Various algorithms for computing the minimizer of the nonconvex
minimization problem (\ref{eq:minimize}) have been studied in
literature. For instance, a continuous normalized gradient flow
(CNGF) and its discretization that preserve the total mass and
magnetization conservation and energy diminishing properties were
presented in \cite{Bao_Wang}.  Perhaps one of the more popular and
efficient techniques for dealing with the normalization
constraints in (\ref{eq:S})  is through the following
construction: choose a time step $k=\Delta t>0$ and denote time
steps as $t_n=n\; k$ for $n=0,1,2,\cdots$\;. To adapt an efficient
algorithm for the solution of the usual gradient flow to the
minimization problem under constraints,  it is natural to consider
the following splitting (or projection) scheme, which was widely
used in the physics literature for computing the ground state of
BECs:
\begin{eqnarray}\label{eq:GFDN_1}
  \p_t\phi_{1}(\bx,t)&=&\left[\fl{1}{2}\nabla^2 -V(\bx)-
(\beta_n+\beta_s)\left(|\phi_1|^2+|\phi_0|^2\right)
-(\beta_n-\beta_s)|\phi_{-1}|^2\right]\phi_1\nn\\
&&-\beta_s\,\bar{\phi}_{-1}\,\phi_0^2, \\
  \label{eq:GFDN_2}
   \p_t\phi_{0}(\bx,t)&=&\left[\fl{1}{2}\nabla^2
-V(\bx)- (\beta_n+\beta_s)\left(|\phi_1|^2+|\phi_{-1}|^2\right)
-\beta_n|\phi_{0}|^2\right]\phi_0 \nn \\
&&-2\beta_s\,\phi_{-1}\,\bar{\phi}_{0}\,\phi_1,\qquad \bx\in{\Bbb
R}^d, \quad t_{n-1}\le t<t_{n},
\qquad n\ge1,\\
  \label{eq:GFDN_3}
\p_t\phi_{-1}(\bx,t)&=&\left[\fl{1}{2}\nabla^2 -V(\bx)-
(\beta_n+\beta_s)\left(|\phi_{-1}|^2+|\phi_0|^2\right)
-(\beta_n-\beta_s)|\phi_{1}|^2\right]\phi_{-1} \nn\\
&&-\beta_s\,\phi_0^2\,\bar{\phi}_{1};
\end{eqnarray}
followed by a projection step as
\begin{eqnarray}\label{eq:projection_1c}
  &&\qquad \phi_1(\bx,t_{n}):=\phi_1(\bx,t_{n}^+)= \sigma_1^{n}\;
  \phi_1(\bx,t_{n}^-), \\
  \label{eq:projection_2c}
  &&\qquad \phi_0(\bx,t_{n}):=\phi_0(\bx,t_{n}^+)= \sigma_0^{n}\;
  \phi_0(\bx,t_{n}^-), \qquad \bx\in{\Bbb
R}^d,\qquad n\ge1,\\
  \label{eq:projection_3c}
  &&\qquad \phi_{-1}(\bx,t_{n}):=\phi_{-1}(\bx,t_{n}^+)=
  \sigma_{-1}^{n}\;
  \phi_{-1}(\bx,t_{n}^-);
\end{eqnarray}
where $\phi_l(\bx,t_n^\pm)=\lim_{t\to t_n^\pm} \phi_l(\bx,t)$
($l=-1,0,1$) and $\sigma_l^{n}$ ($l=-1,0,1$) are projection
constants and they are chosen such that \be\label{cong1}
\|\Phi(\cdot,t_{n})\|^2=\sum_{l=-1}^1\|\phi_l(\cdot,t_{n})\|^2
=1,\qquad   \|\phi_1(\cdot,t_{n})\|^2-
\|\phi_{-1}(\cdot,t_{n})\|^2=M. \ee In fact, the gradient flow
(\ref{eq:GFDN_1})-(\ref{eq:GFDN_3}) can be viewed as applying the
steepest decent method to the energy functional $E(\Phi)$ in
(\ref{energy}) without constraints, and
(\ref{eq:projection_1c})-(\ref{eq:projection_3c}) project the
solution back to the unit sphere $S$ in order to satisfy the
constraints in (\ref{eq:S}). In addition,
(\ref{eq:GFDN_1})-(\ref{eq:GFDN_3}) can also be obtained from the
coupled GPEs (\ref{eq:dCGPE_1})-(\ref{eq:dCGPE_3}) by the change
of variable $t\to -i\; t$, that is why the algorithm is usually
called as the imaginary time method in the physics literatures
\cite{Chiofalo,Bao_Du,Bao3}.

Plugging (\ref{eq:projection_1c})-(\ref{eq:projection_3c}) into
(\ref{cong1}), we obtain \bea\label{cong2} &&\sum_{l=-1}^1
\left(\sigma_l^{n}\right)^2
  \|\phi_l(\cdot,t_{n}^-)\|^2=1,\\
  \label{cong3}
  &&\left(\sigma_1^{n}\right)^2
  \|\phi_1(\cdot,t_{n}^-)\|^2-\left(\sigma_{-1}^{n}\right)^2
  \|\phi_{-1}(\cdot,t_{n}^-)\|^2=M.
  \eea
There are three unknowns and only two equations in the above nonlinear
system, so the solution is undetermined! In order to determine
the projection constants $\sigma_l^{n}$ ($l=-1,0,1$), we
need to find an additional equation. Based on the relation
between the chemical potentials in (\ref{chem8}) and the
continuous normalized gradient flow proposed in \cite{Bao_Wang}
for computing the ground state of spin-1 BEC, see details in
Appendix A, we propose to use the following equation as the
third normalization condition \be\label{cong5}
\sigma_1^{n}\; \sigma_{-1}^{n}=\left(\sigma_0^{n}\right)^2. \ee
Solving the nonlinear system (\ref{cong2}), (\ref{cong3}) and
(\ref{cong5}), see details in Appendix B, we get explicitly the
projection constants as \be\label{eq:constant1}
  \sigma_0^{n}=\frac{\sqrt{1-M^2}}
  {\left[\|\phi_0(\cdot,t_{n}^-)\|^2 +
  \sqrt{4(1-M^2)\|\phi_1(\cdot,t_{n}^-)\|^2
  \|\phi_{-1}(\cdot,t_{n}^-)\|^2
  + M^2\|\phi_0(\cdot,t_{n}^-)\|^4}\right]^{1/2}},
  \ee
  \be
  \label{eq:constant2}
  \sigma_1^{n}=\frac{\sqrt{1+M-(\sg_0^n)^2
  \|\phi_0(\cdot,t_{n}^-)\|^2}}
  {\sqrt{2}\ \|\phi_1(\cdot,t_{n}^-)\|}, \qquad
  \sigma_{-1}^{n}=\frac{\sqrt{1-M-(\sg_0^n)^2
  \|\phi_0(\cdot,t_{n}^-)\|^2}}
  {\sqrt{2}\ \|\phi_{-1}(\cdot,t_{n}^-)\|}.
 \ee
From the numerical point of view, the gradient flow
(\ref{eq:GFDN_1})-(\ref{eq:GFDN_3}) can be solved via traditional
techniques, and the normalization of the gradient flow
is simply achieved by a projection at the end of each time step.

\section{Backward-forward Euler sine-pseudospectral method}
\label{s3}

 In this section, we will present the backward-forward
Euler sine-pseudospectral method (BESP) to discretize the
normalized gradient flow (\ref{eq:GFDN_1})-(\ref{eq:GFDN_3}),
(\ref{eq:projection_1c})-(\ref{eq:projection_3c}) and
(\ref{eq:constant1})-(\ref{eq:constant2}).

 Due to the trapping potential $V(\bx)$ given by (\ref{poten}),
the solution $\Phi(\bx,t)$  decays to zero exponentially fast when
$|\bx|\to\infty$. Thus in practical computation, we truncate the
problem into a bounded computational domain $\Omega_{\bx}$ (chosen
as an interval $(a,b)$ in 1D, a rectangle $(a,b)\times (c,d)$ in
2D, and a box $(a,b)\times (c,d)\times (e,f)$ in 3D, with $|a|$,
$|c|$, $|e|$, $b$, $d$ and $f$
 sufficiently large) with homogeneous
Dirichlet boundary conditions.

For simplicity of notation we introduce the method for the case of
one spatial dimension ($d=1$) defined over the interval $(a,b)$
with homogeneous Dirichlet boundary conditions. Generalization to
higher dimension are straightforward for tensor product grids, and
the results remain valid without modifications. For $d=1$,
we choose the spatial mesh size $h=\Delta x>0$ with $h=(b-a)/M$
for $M$ an even positive integer, and let the grid points be
\[x_l:= a + j \; h, \qquad j=0,1,\cdots, M.\]
Let $\Phi_j^n=(\phi_{1,j}^n,\phi_{0,j}^n,\phi_{-1,j}^n)^T$ be the
approximation of $\Phi(x_j,t_n)=(\phi_1(x_j,t_n)$,
$\phi_0(x_j,t_n)$, $\phi_{-1}(x_j,t_n))^T$ and $\Phi^n$ be the
solution vector with component $\Phi_j^n$. In the discretization,
we use sine-pseudospectral method for spatial derivatives and
backward/forward  Euler scheme for linear/nonlinear terms in time
discretization.
 The gradient flow (\ref{eq:GFDN_1})-(\ref{eq:GFDN_3}) is
 discretized, for $j=1,2,\ldots,M-1$ and $n\ge1$,  as
\bea\label{eq:space_1}
&&\frac{\phi_{1,j}^*-\phi_{1,j}^{n-1}}{\Delta t}
  = \frac{1}{2}D^s_{xx}\phi_1^*|_{x=x_j} - \alpha_1 \phi_{1,j}^*
  + G_{1,j}^{n-1}, \\
\label{eq:space_2} &&\frac{\phi_{0,j}^*-\phi_{0,j}^{n-1}}{\Delta
t}
  = \frac{1}{2}D^s_{xx}\phi_0^*|_{x=x_j} - \alpha_0 \phi_{0,j}^*
  + G_{0,j}^{n-1}, \qquad ,\\
\label{eq:space_3}
&&\frac{\phi_{-1,j}^*-\phi_{-1,j}^{n-1}}{\Delta
t}
  = \frac{1}{2}D^s_{xx}\phi_{-1}^*|_{x=x_j} - \alpha_{-1} \phi_{-1,j}^*
  + G_{-1,j}^{n-1};
 \eea
where \bea
G_{1,j}^{n-1}&=&\left[\ap_1-V(x_j)-(\beta_n+\beta_s)
\left(|\phi_{1,j}^{n-1}|^2+|\phi_{0,j}^{n-1}|^2\right)
-(\beta_n-\beta_s)|\phi_{-1,j}^{n-1}|^2\right]\phi_{1,j}^{n-1}\nn\\
&&-\beta_s\,\bar{\phi}_{-1,j}^{n-1}\,\left(\phi_{0,j}^{n-1}\right)^2,\\
G_{0,j}^{n-1}&=&\left[\ap_0-V(x_j)-
(\beta_n+\beta_s)\left(|\phi_{1,j}^{n-1}|^2+|\phi_{-1,j}^{n-1}|^2\right)
-\beta_n|\phi_{0,j}^{n-1}|^2\right]\phi_{0,j}^{n-1}\nn\\
&&-2\beta_s\,\phi_{-1,j}^{n-1}\,\bar{\phi}_{0,j}^{n-1}\,\phi_{1,j}^{n-1},\\
G_{-1,j}^{n-1}&=&\left[\ap_{-1}-V(x_j)-
(\beta_n+\beta_s)\left(|\phi_{-1,j}^{n-1}|^2+|\phi_{0,j}^{n-1}|^2\right)
-(\beta_n-\beta_s)|\phi_{1,j}^{n-1}|^2\right]\phi_{-1,j}^{n-1}\nn\\
&&-\beta_s\,\left(\phi_{0,j}^{n-1}\right)^2\,\bar{\phi}_{1,j}^{n-1}.
\eea Here, $D_{xx}^s$, a pseudospectral differential operator
approximation of $\partial_{xx}$, is defined as
\[\left. D_{xx}^s U\right|_{x=x_j}= -\sum_{m=1}^{M-1}\;
\mu_m^2 (\hat{U})_m\; \sin(\mu_m(x_j-a)), \qquad j=1,2,\cdots,M-1,
\] where $(\hat{U})_m$ ($m=1,2,\cdots,M-1$), the sine transform
coefficients of the vector $U=(U_0,U_1,\cdots,U_M)^T$ satisfying
$U_0=U_M=0$, are defined as
\[\mu_m=\frac{\pi m}{b-a}, \qquad
(\hat{U})_m=\frac{2}{M}\sum_{j=1}^{M-1} \; U_j\;
\sin(\mu_m(x_j-a)), \quad m=1,2,\cdots,M-1;\] and $\ap_l$
($l=-1,0,1$) are the stabilization parameters which are chosen in
the `optimal' form (such that the time step
can be chosen as large as possible)
as \cite{Bao_Chern_Lim} \be \qquad \alpha_1 =
\frac{1}{2}\left(b_1^{\rm max} + b_1^{\rm min}\right), \quad
\alpha_0 = \frac{1}{2}\left(b_0^{\rm max} + b_0^{\rm min}\right),
\quad \alpha_{-1} = \frac{1}{2}\left(b_{-1}^{\rm max} +
b_{-1}^{\rm min}\right); \ee with \beas &&b_1^{\rm max}=
\max_{1\le j \le
M-1}\left[V(x_j)+(\beta_n+\beta_s)\left(|\phi_{1,j}^{n-1}|^2
+|\phi_{0,j}^{n-1}|^2\right)
+(\beta_n-\beta_s)|\phi_{-1,j}^{n-1}|^2\right], \\
&&b_1^{\rm min}= \min_{1\le j \le
M-1}\left[V(x_j)+(\beta_n+\beta_s)\left(|\phi_{1,j}^{n-1}|^2
+|\phi_{0,j}^{n-1}|^2\right)
+(\beta_n-\beta_s)|\phi_{-1,j}^{n-1}|^2\right], \\
&&b_0^{\rm max}=\max_{1\le j \le M-1}\left[V(x_j)+
(\beta_n+\beta_s)\left(|\phi_{1,j}^{n-1}|^2+|\phi_{-1,j}^{n-1}|^2\right)
+\beta_n|\phi_{0,j}^{n-1}|^2\right],\\
&&b_0^{\rm min}=\min_{1\le j \le M-1}\left[V(x_j)+
(\beta_n+\beta_s)\left(|\phi_{1,j}^{n-1}|^2+|\phi_{-1,j}^{n-1}|^2\right)
+\beta_n|\phi_{0,j}^{n-1}|^2\right],\\
&&b_{-1}^{\rm max}=\max_{1\le j \le M-1}\left[V(x_j)+
(\beta_n+\beta_s)\left(|\phi_{-1,j}^{n-1}|^2+|\phi_{0,j}^{n-1}|^2\right)
+(\beta_n-\beta_s)|\phi_{1,j}^{n-1}|^2\right],\\
&&b_{-1}^{\rm min}=\min_{1\le j \le M-1}\left[V(x_j)+
(\beta_n+\beta_s)\left(|\phi_{-1,j}^{n-1}|^2+|\phi_{0,j}^{n-1}|^2\right)
+(\beta_n-\beta_s)|\phi_{1,j}^{n-1}|^2\right]. \eeas The
homogeneous Dirichlet boundary conditions are discretized as
 \be\label{bondd}
\phi_{1,0}^*=\phi_{1,M}^*=\phi_{0,0}^*=\phi_{0,M}^*
=\phi_{-1,0}^*=\phi_{-1,M}^*=0. \ee The projection step
(\ref{eq:projection_1c})-(\ref{eq:projection_1c}) is discretized,
for $0\le j\le M$ and $n\ge1$, as \bea\label{proj1}
&&\phi_{1,j}^n= \sigma_1^{n}\;
  \phi_{1,j}^*, \quad
  \phi_{0,j}^n= \sigma_0^{n}\;
  \phi_{0,j}^*,
  \quad \phi_{-1,j}^n= \sigma_{-1}^{n}\;
  \phi_{-1,j}^*,
\eea where \be\label{proj4}
  \sigma_0^{n}=\frac{\sqrt{1-M^2}}
  {\left[\|\phi_0^*\|^2 +
  \sqrt{4(1-M^2)\|\phi_1^*\|^2
  \|\phi_{-1}^*\|^2
  + M^2\|\phi_0^*\|^4}\right]^{1/2}},
  \ee
  \be
  \label{proj5}
  \sigma_1^{n}=\frac{\sqrt{1+M-\alpha_0^2
  \|\phi_0^*\|^2}}
  {\sqrt{2}\ \|\phi_1^*\|}, \qquad
  \sigma_{-1}^{n}=\frac{\sqrt{1-M-\alpha_0^2
  \|\phi_0^*\|^2}}
  {\sqrt{2}\ \|\phi_{-1}^*\|};
 \ee
with
\[\|\phi_1^*\|^2=h\sum_{j=1}^{M-1} |\phi_{1,j}^*|^2, \qquad
\|\phi_0^*\|^2=h\sum_{j=1}^{M-1} |\phi_{0,j}^*|^2,\qquad
\|\phi_{-1}^*\|^2=h\sum_{j=1}^{M-1} |\phi_{-1,j}^*|^2. \] The
initial data (\ref{init}) is discretized as
\[\phi_{l,j}^0=\phi_l(x_j,0), \qquad j=0,1,2,\cdots,M, \qquad
l=-1,0,1.
\]

The linear system (\ref{eq:space_1})-(\ref{eq:space_3}) can be
solved very efficiently by using the fast sine transform. In fact,
take discrete sine transform at both sides, we get \bea
\label{dstt1} &&\qquad \frac{1}{\Delta
t}\left[(\hat{\phi}_1^*)_m-(\hat{\phi}_1^{n-1})_m\right]=-\left[\frac{1}{2}
\mu_m^2 +\ap_1\right]
(\hat{\phi}_1^*)_m+(\hat{G}_1^{n-1})_m,\\
\label{dstt2} &&\qquad \frac{1}{\Delta
t}\left[(\hat{\phi}_0^*)_m-(\hat{\phi}_0^{n-1})_m\right]=-\left[\frac{1}{2}
\mu_m^2+\ap_0\right]
(\hat{\phi}_0^*)_m+(\hat{G}_0^{n-1})_m,\quad 1\le m<M,\\
\label{dstt3} &&\qquad \frac{1}{\Delta
t}\left[(\hat{\phi}_{-1}^*)_m-(\hat{\phi}_{-1}^{n-1})_m\right]
=-\left[\frac{1}{2}
\mu_m^2+ \ap_{-1}\right]
(\hat{\phi}_{-1}^*)_m+(\hat{G}_{-1}^{n-1})_m. \eea Solve the above
system in the phase space, we obtain \bea \label{dstt4}
&&(\hat{\phi}_1^*)_m=\frac{1}{1+\Delta
t\left[\ap_1+\mu_m^2/2\right]}\left[(\hat{\phi}_1^{n-1})_m
+(\hat{G}_1^{n-1})_m\right],\\
\label{dstt5} &&(\hat{\phi}_0^*)_m=\frac{1}{1+\Delta
t\left[\ap_1+\mu_m^2/2\right]}\left[(\hat{\phi}_0^{n-1})_m
+(\hat{G}_0^{n-1})_m\right],\quad 1\le m<M,\\
\label{dstt6} &&(\hat{\phi}_{-1}^*)_m=\frac{1}{1+\Delta
t\left[\ap_1+\mu_m^2/2\right]}\left[(\hat{\phi}_{-1}^{n-1})_m
+(\hat{G}_{-1}^{n-1})_m\right]. \eea

\begin{remark} The gradient flow (\ref{eq:GFDN_1})-(\ref{eq:GFDN_3})
can also be discretized by using the backward Euler finite
difference method proposed in \cite{Bao_Du} or the backward Euler
sine-pseudospectral method proposed in \cite{Bao_Chern_Lim} for
computing the ground state of  one-component BEC.
\end{remark}

\section{Numerical Results}
\label{s4} \setcounter{equation}{0}

In this section, we first show that the ground states computed by
our new numerical method are independent of the  choice of the
initial data in (\ref{init})  and verify numerically the energy
diminishing property of the method. Finally, we apply
the method to  compute the ground state of spin-1  BEC with
different interactions and trapping potentials.
 In our computations, the ground state is
reached by using the numerical method
(\ref{eq:space_1})-(\ref{eq:space_3}), (\ref{proj1})-(\ref{proj5})
when $\|\Phi_h^{n+1}-\Phi_h^n\|\le \vep:=10^{-7}$.
In addition, in the ground state of spin-1 BEC, we have
$M\leftrightarrow -M\ \Longleftrightarrow\ \phi_1 \leftrightarrow
\phi_{-1}$, thus we only present results for $0\le M\le 1$.


\subsection{Choice of initial data}

In our tests, two typical physical experiments are considered:

\begin{itemize}

\item  Case I. With ferromagnetic interaction, e.g. $^{87}$Rb
confined in a cigar-shaped trapping potential with parameters: $m
= 1.443 \times 10^{-25}$[kg], $a_0 = 5.387$[nm], $a_2 =
5.313$[nm], $\omega_x = 2\pi \times 20$[Hz], $\omega_y = \omega_z
= 2\pi \times 400$[Hz]. This suggests us to use dimensionless
quantities in (\ref{eq:dCGPE_1})-(\ref{eq:dCGPE_3}) for our
computations as: $d=1$, $V(x)=x^2/2$, $\beta_n\approx
\frac{4\pi(a_0+2a_2)N}{3a_s}\frac{\sqrt{\og_y\og_z}}{2\pi\og_x}=0.0885N$
 and $\beta_s \approx
\frac{4\pi(a_2-a_0)N}{3a_s}\frac{\sqrt{\og_y\og_z}}{2\pi\og_x}
=-0.00041N$ with $N$ the total number of atoms in the condensate
and the dimensionless length unit
$a_s=\sqrt{\hbar/m\og_x}=2.4116\times 10^{-6}$ [m] and time unit
$t_s=1/\og_x=0.007958$[s].

\item Case II. With antiferromagnetic interaction, e.g. $^{23}$Na
confined in a cigar-shaped trapping potential with parameters: $m
= 3.816 \times 10^{-26}$[kg], $a_0 = 2.646$[nm], $a_2 =
2.911$[nm], $\omega_x = 2\pi \times 20$[Hz], $\omega_y = \omega_z
= 2\pi \times 400$[Hz]. Again, this suggests us to use
the following dimensionless quantities
in our computations: $d=1$, $V(x)=x^2/2$, $\beta_n\approx 0.0241N$
 and $\beta_s \approx 0.00075N$ with the dimensionless length unit
 $a_s=4.6896\times 10^{-6}$ [m]
and time unit $t_s=0.007958$[s].
\end{itemize}

  We first test that the converged solution is independent of
different choices of the initial data in (\ref{init}) and energy
diminishing property of the normalized gradient flow. In order to
do so, we choose the initial data in (\ref{init}) as
\begin{itemize}
\item Gaussian profiles satisfying the constraints in (\ref{eq:S})
initially, i.e.
\begin{eqnarray}
\label{initf1}
\phi_{1}(x,0) &=& \frac{\sqrt{0.5(1+M-\kappa)}}{\pi^{1/4}}e^{-x^2/2}, \\
\label{initf2}
\phi_0(x,0) &=& \frac{\sqrt{\kappa}}{\pi^{1/4}}e^{-x^2/2},
 \qquad -\ift<x<\ift, \\
\label{initf3} \phi_{-1}(x,0) &=&
\frac{\sqrt{0.5(1-M-\kappa)}}{\pi^{1/4}}e^{-x^2/2},
    \end{eqnarray}
where $\kappa$ is a constant satisfying $0<\kappa<1-|M|$.

\item Unnormalized Gaussian profiles, i.e.
\begin{equation}\label{initf6}
\phi_1(x,0) = \phi_0(x,0) = \phi_{-1}(x,0) =e^{-x^2/2},
 \qquad \ift<x<\ift.
\end{equation}
\end{itemize}

\begin{figure}[htb]
\centerline{\psfig{figure=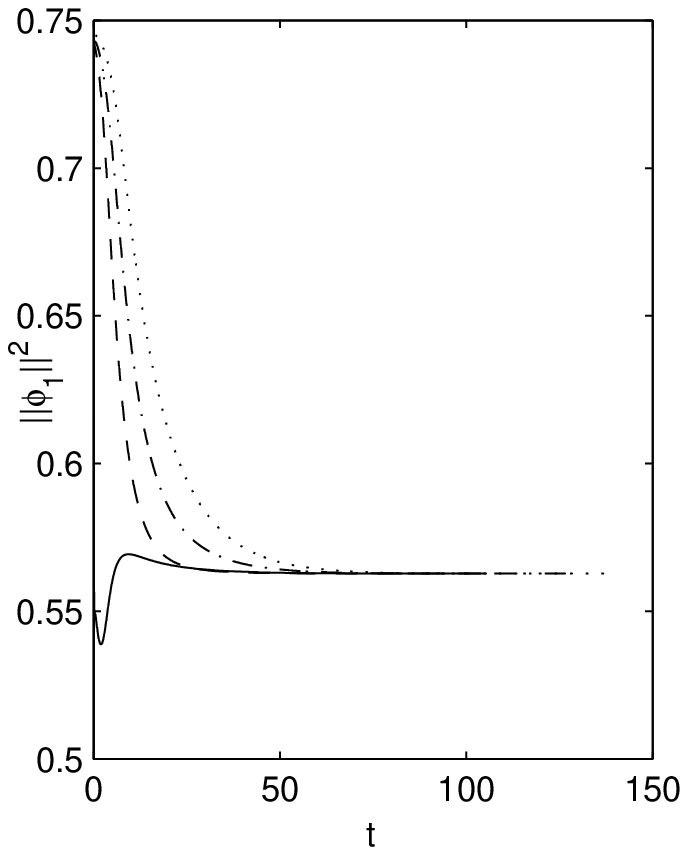,height=4.5cm,width=3.8cm,angle=0}
\quad
\psfig{figure=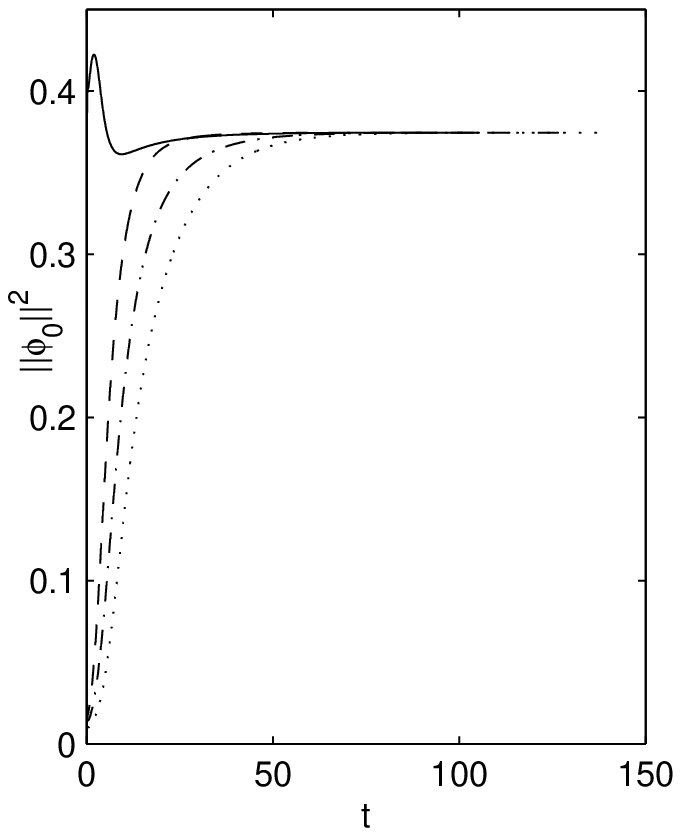,height=4.5cm,width=3.8cm,angle=0}
\quad
\psfig{figure=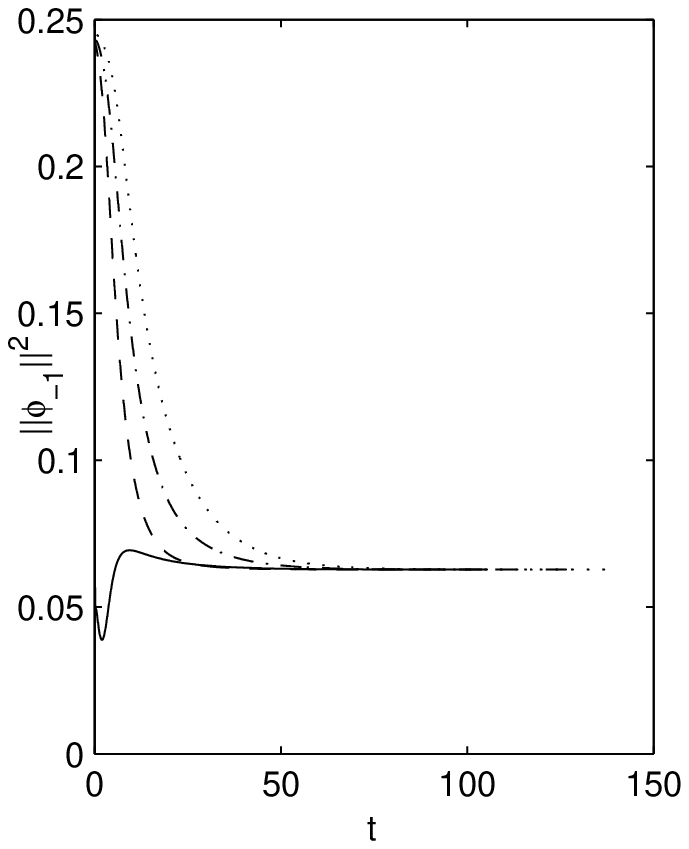,height=4.5cm,width=3.8cm,angle=0} }
\vspace{0.2cm} \caption{Time evolution of
$N_1=\|\phi_1(\cdot,t)\|^2$ (`left'),
$N_0=\|\phi_0(\cdot,t)\|^2$ (`middle') and
$N_{-1}=\|\phi_{-1}(\cdot,t)\|^2$ (`right') by our method
(\ref{eq:projection_1c})-(\ref{eq:projection_3c}) for $^{87}$Rb in
Case I with $M=0.5$ and $N=10^4$ to analyze the
convergence of different initial data in (\ref{initf6}) (solid
line) and (\ref{initf1})-(\ref{initf3}) with $\kappa=0.1$ (dotted
line), $\kappa=0.2$ (dash-dot line) and  $\kappa=0.4$ (dashed
line), respectively.  } \label{fig:converge_Rb}
\end{figure}

We solve the problem (\ref{eq:minimize}) by our method on
$[-16,16]$  with time step $\btu t=0.005$ and mesh size $h=1/64$
for different values of $\kappa$ in (\ref{initf1})-(\ref{initf3}).
 Figure \ref{fig:converge_Rb} plots time evolution of
$N_l(t):=\|\phi_l(\cdot,t)\|^2$ ($l=1,0,-1$) for $^{87}$Rb in Case
I with $M=0.5$ and $N=10^4$ for different choices
of the initial data in (\ref{initf6}) and
(\ref{initf1})-(\ref{initf3}), and Figure \ref{fig:converge_Na} shows
similar results for $^{23}$Na in Case II.
In addition, Figure \ref{fig:energy} depicts time evolution of
the energy for the two cases with $M=0.5$ and $N=10^4$ for different choices
of the initial data in (\ref{initf6}).

\begin{figure}[htb]
\centerline{\psfig{figure=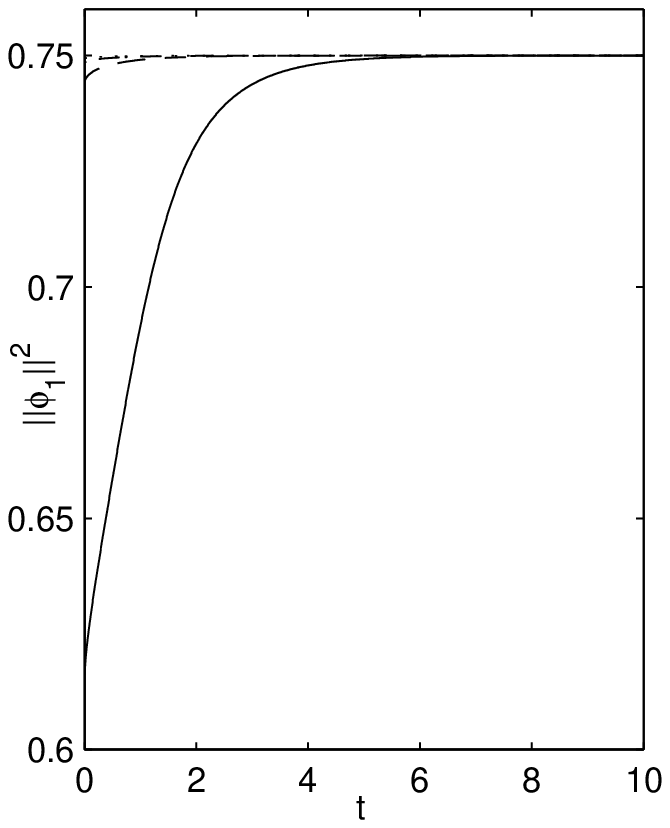,height=4.5cm,width=3.8cm,angle=0}
\quad
\psfig{figure=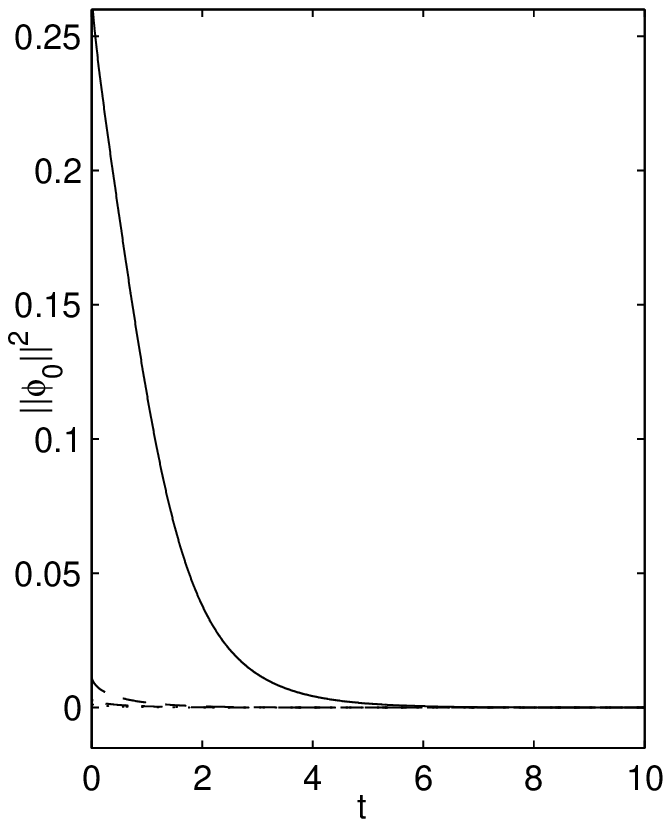,height=4.5cm,width=3.8cm,angle=0}
\quad
\psfig{figure=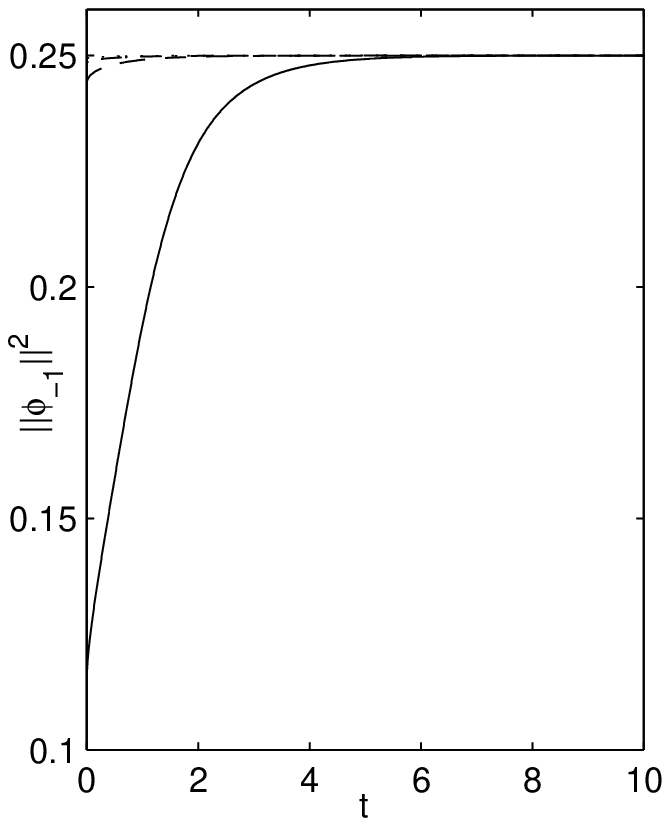,height=4.5cm,width=3.8cm,angle=0} }
\vspace{0.2cm}

\caption{Time evolution of $N_1=\|\phi_1(\cdot,t)\|^2$ (`left'),
$N_0=\|\phi_0(\cdot,t)\|^2$ (`middle')
and $N_{-1}=\|\phi_{-1}(\cdot,t)\|^2$ (`right') by our method
(\ref{eq:projection_1c})-(\ref{eq:projection_3c}) for $^{23}$Na in
Case II with $M=0.5$, and $N=10^4$ to analyze the
convergence of different initial data in (\ref{initf6}) (solid
line) and (\ref{initf1})-(\ref{initf3}) with $\kappa=0.1$ (dotted
line), $\kappa=0.2$ (dash-dot line) and  $\kappa=0.4$ (dashed
line), respectively. } \label{fig:converge_Na}
\end{figure}

\begin{figure}[htb]
\centerline{a)\psfig{figure=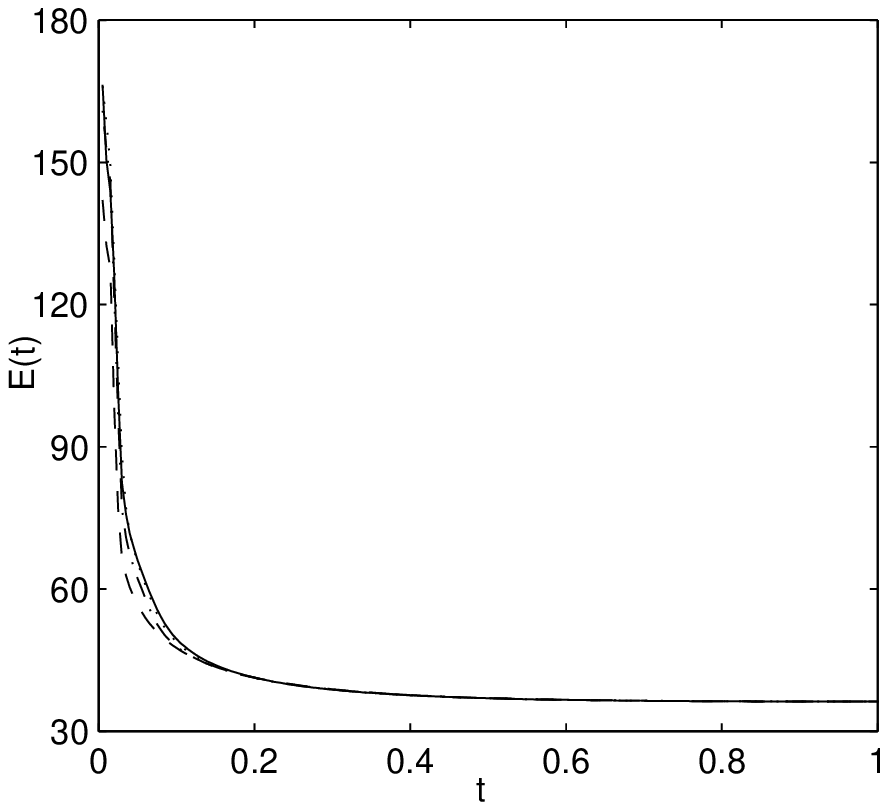,height=4.5cm,width=6cm,angle=0}
\quad
b)\psfig{figure=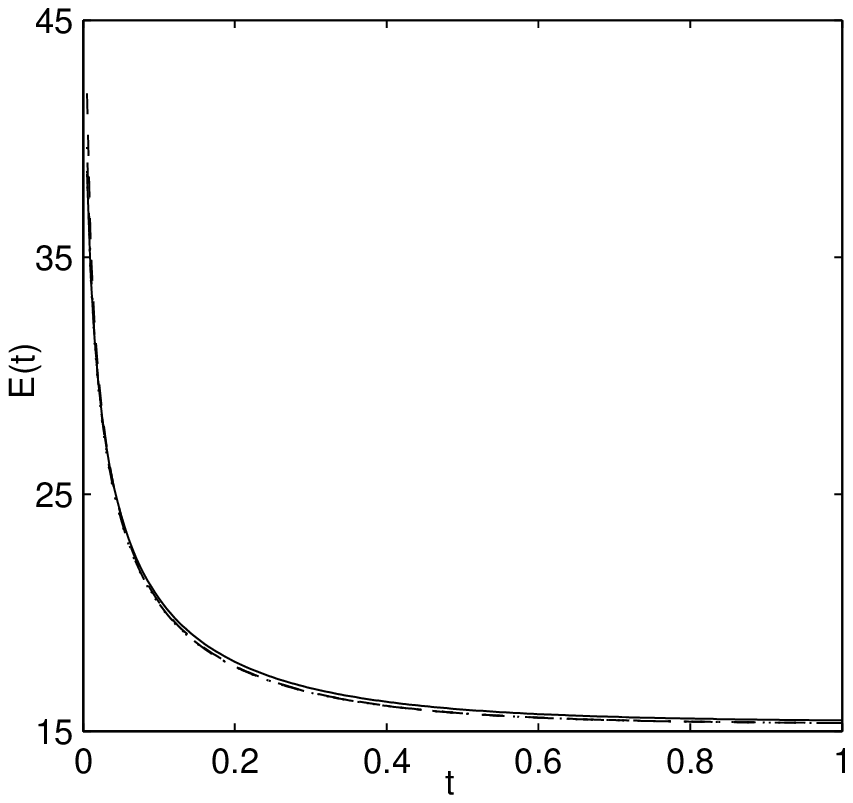,height=4.5cm,width=6cm,angle=0}}
\caption{Time evolution of the energy by our method
(\ref{eq:projection_1c})-(\ref{eq:projection_3c})
 with $M=0.5$, and $N=10^4$  for
 a) $^{87}$Rb in case I; and b) $^{23}$Na in case II with
 different initial data in (\ref{initf6}) (solid
line) and (\ref{initf1})-(\ref{initf3}) with $\kappa=0.1$ (dotted
line), $\kappa=0.2$ (dash-dot line) and  $\kappa=0.4$ (dashed
line), respectively. }
\label{fig:energy}
\end{figure}

From Figs. \ref{fig:converge_Rb} and \ref{fig:converge_Na},
we can
see that the converged ground states are independent of the choice
of initial data. In fact, based on our extensive numerical
experiments on other types of initial data
(not shown here for brevity), our
numerical method always gives the ground state if all the three
components in the initial data are chosen as nonnegative functions.
In addition, Fig. \ref{fig:energy} demonstrates the energy diminishing
property of the normalized gradient flow and its full discretization
when time step $\Delta t$ is small.
Based on our numerical experiments, for $0\le M\le 1$,
we suggest the initial data
in (\ref{init}) be chosen as: i)
with ferromagnetic interaction, i.e. $\beta_s\le0$
\[
\phi_1(\bx)=\frac{1}{2}\sqrt{1+3M}\phi_g^{\rm ap}(\bx),
\quad \phi_0(\bx)=\sqrt{\frac{1-M}{2}}\phi_g^{\rm ap}(\bx),
\quad \phi_1(\bx)=\frac{1}{2}\sqrt{1-M}\phi_g^{\rm ap}(\bx);
\]
and ii) with antiferromagnetic interaction, i.e. $\beta_s>0$
\[
\phi_1(\bx)=\sqrt{\frac{1+M}{2}}\phi_g^{\rm ap}(\bx),
\quad \phi_0(\bx)=0,
\quad \phi_1(\bx)=\sqrt{\frac{1-M}{2}}\phi_g^{\rm ap}(\bx);
\]
where $\phi_g^{\rm ap}(\bx)$ can be chosen as
the approximate ground state solution of single component
BEC, e.g. the harmonic oscillator approximation when $\beta_n$ small
and the Thomas-Fermi approximation when $\beta_n\gg1$
\cite{Bao_Du, BaoT, BaoLZ}.
Based on these choices of initial data,
we report the ground states computed by our numerical method.

Figure \ref{fig:1} shows the ground state solutions of $^{87}$Rb
in Case I with $N=10^4$ for
different magnetization $M$ and Table \ref{tbl:1} lists the
corresponding ground state energies and their Lagrange multipliers
(see their detailed formulation in Appendix C). In addition,
Figure \ref{fig:2} shows similar ground state solutions with
$M=0.5$ for different particle number $N$.

\begin{figure}[htb]
\centerline{\psfig{figure=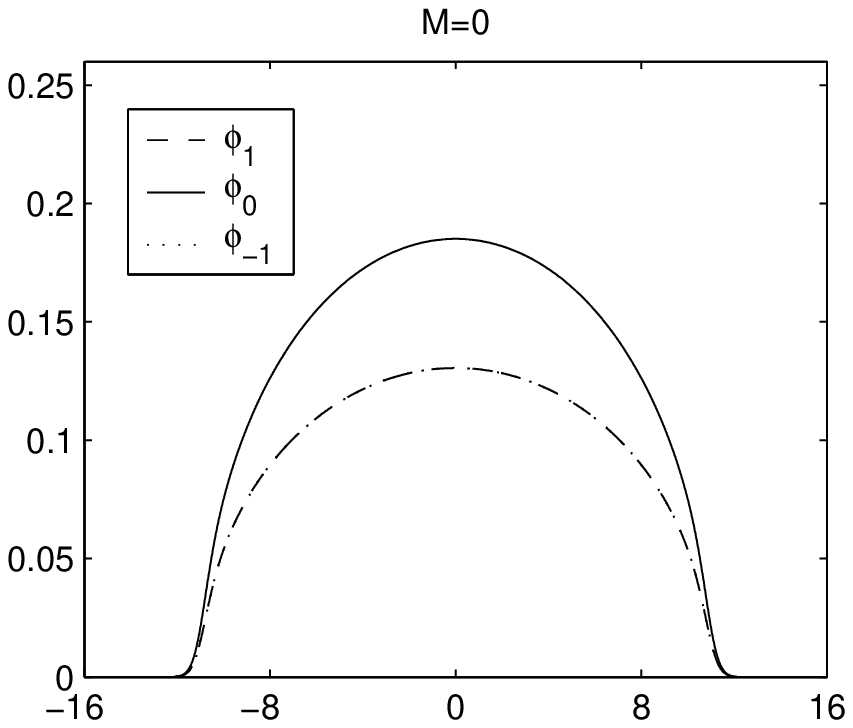,height=5cm,width=5.5cm,angle=0}
\quad \psfig{figure=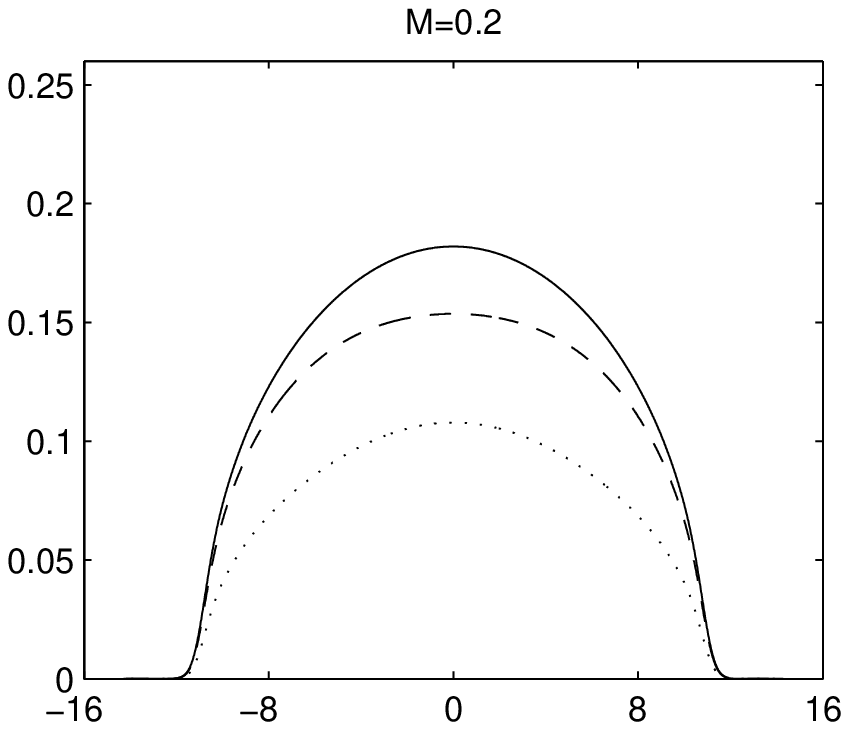,height=5cm,width=5.5cm,angle=0}}
\centerline{\psfig{figure=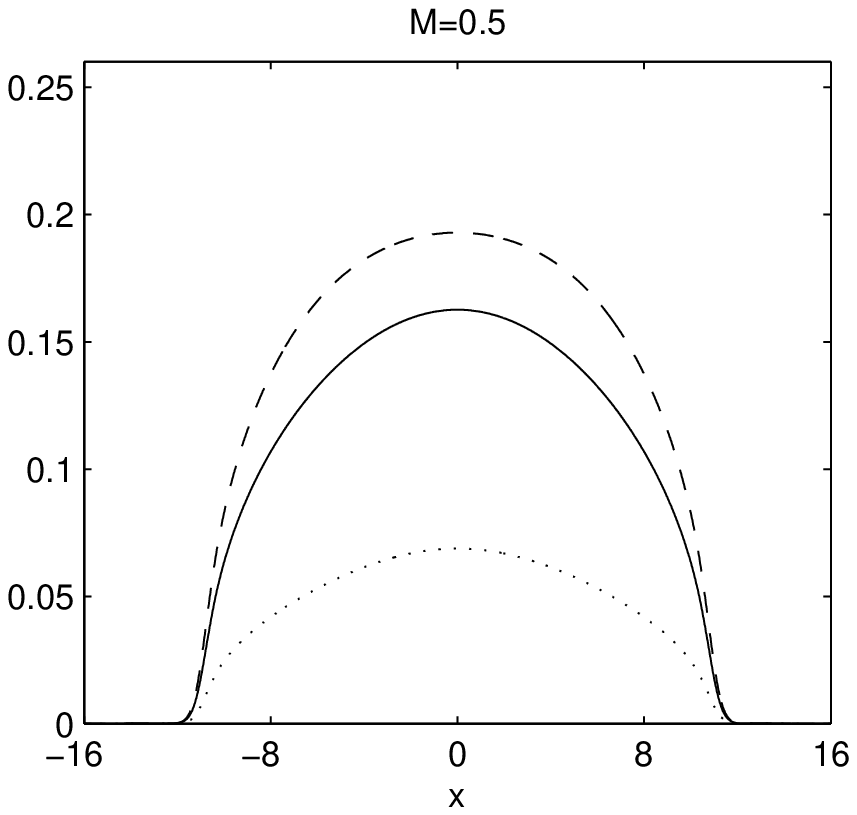,height=5.2cm,width=5.5cm,angle=0}
\quad \psfig{figure=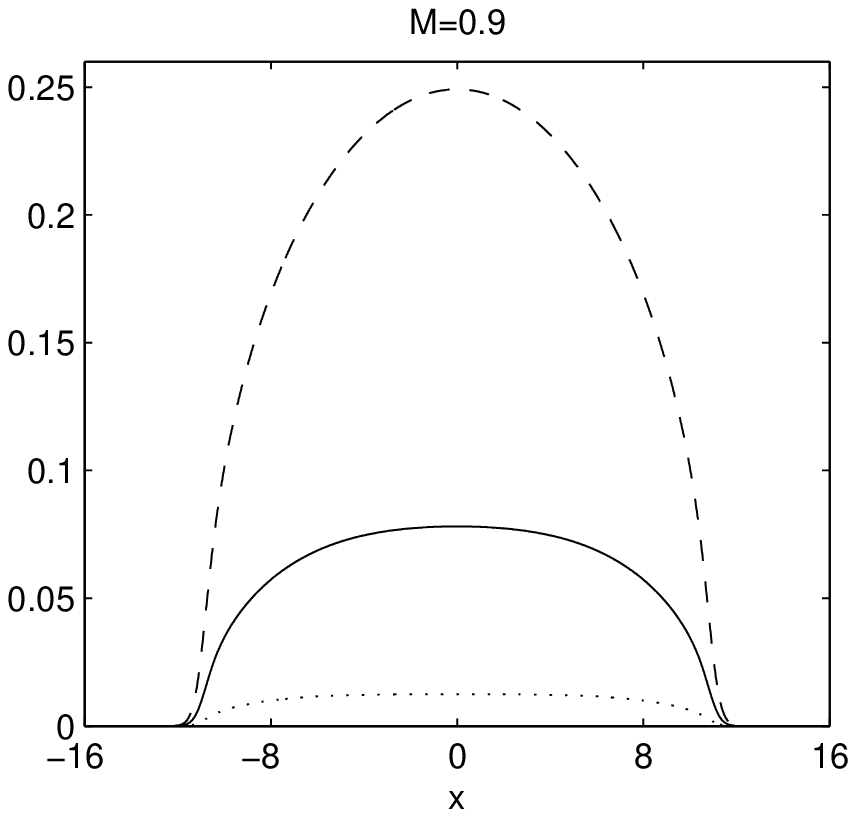,height=5.2cm,width=5.5cm,angle=0}}
\vspace{0.1cm}

\caption{Wave functions of the ground state, i.e. $\phi_1(x)$
(dashed line), $\phi_0(x)$ (solid line)
and $\phi_{-1}(x)$ (dotted  line), of
$^{87}$Rb in Case I  with fixed number of particles $N=10^4$  for
different magnetization $M=0,0.2,0.5,0.9$. } \label{fig:1}
\end{figure}

\begin{table}[htbp]
\begin{center}
\begin{tabular}{cccc}    \hline
 $M$  & $E$  & $\mu$  & $\lambda (\times 10^{-5})$   \\   \hline
 0 & 36.1365 & 60.2139 & 0  \\
 0.1 & 36.1365 & 60.2139 & 1.574  \\
 0.2 & 36.1365 & 60.2139 & 1.621  \\
 0.3 & 36.1365 & 60.2139 & 1.702  \\
 0.4 & 36.1365 & 60.2139 & 1.827  \\
 0.5 & 36.1365 & 60.2139 & 2.014  \\
 0.6 & 36.1365 & 60.2139 & 2.218  \\
 0.7 & 36.1365 & 60.2139 & 2.062  \\
 0.8 & 36.1365 & 60.2139 & 2.081  \\
 0.9 & 36.1365 & 60.2139 & 2.521 \\
 \hline
\end{tabular}
\end{center}
\caption{Ground state energy $E$ and their chemical potentials
$\mu$ and $\lambda$  for $^{87}$Rb in Case I  with $N=10^4$ for
different magnetization $M$.} \label{tbl:1}
\end{table}


\begin{figure}[htb]
\centerline{\psfig{figure=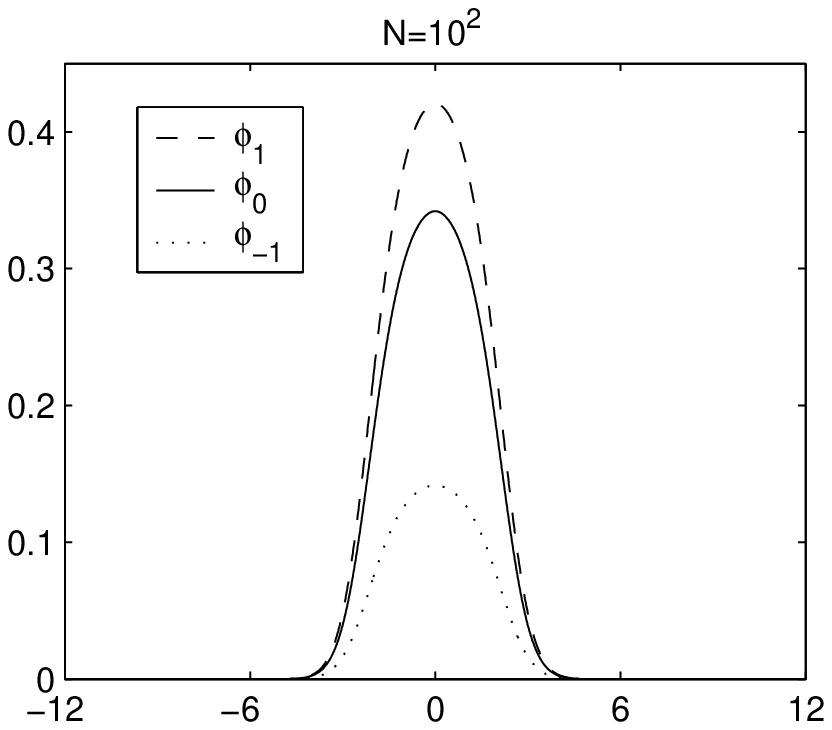,height=5cm,width=5.5cm,angle=0}
\quad \psfig{figure=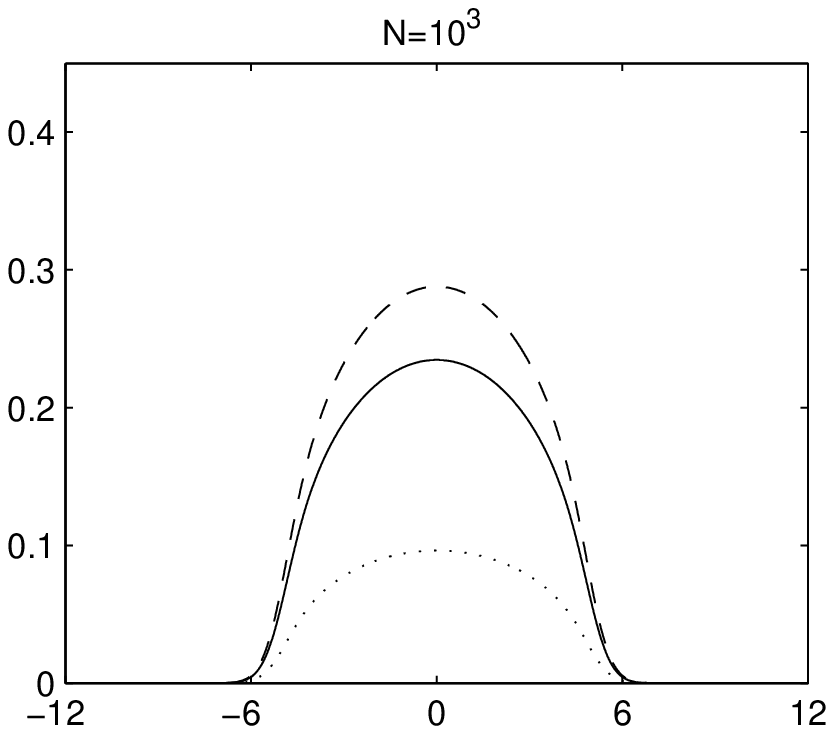,height=5cm,width=5.5cm,angle=0}}
\centerline{\psfig{figure=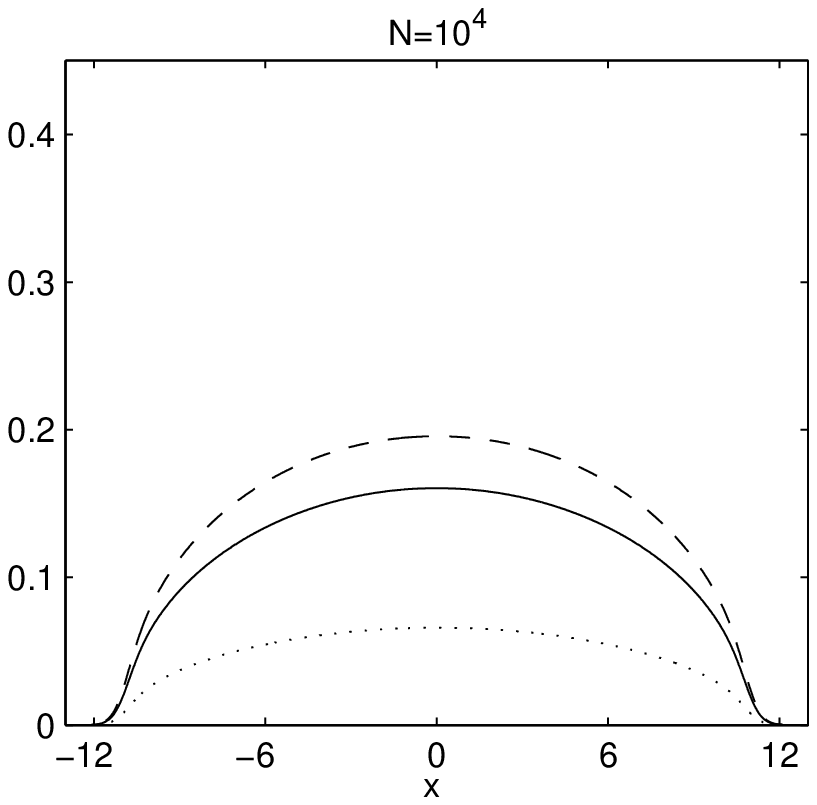,height=5.2cm,width=5.5cm,angle=0}
\quad
\psfig{figure=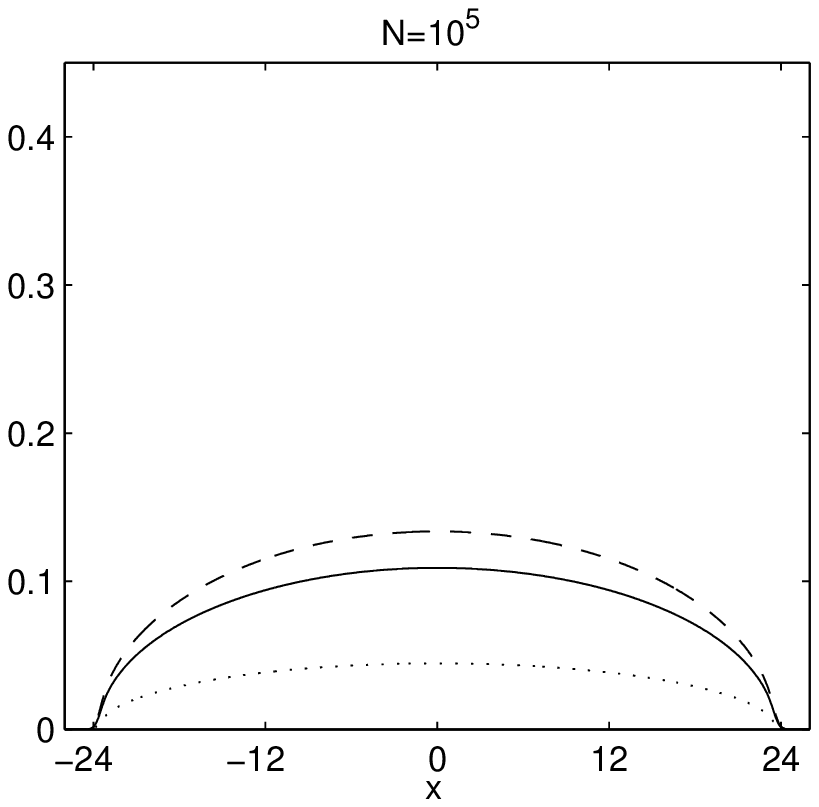,height=5.2cm,width=5.5cm,angle=0}}
\vspace{0.1cm}

\caption{Wave functions of the ground state, i.e. $\phi_1(x)$
(dashed line), $\phi_0(x)$ (solid line) and $\phi_{-1}(x)$ (dotted
line), of $^{87}$Rb in Case I with magnetization $M=0.5$  for
different number of particles $N$. } \label{fig:2}
\end{figure}

Similarly, Figure \ref{fig:5} shows the ground state solutions of
$^{23}$Na in Case II with $N=10^4$
for different magnetization $M$ and Table \ref{tbl:2} lists the
corresponding ground state energies and their Lagrange
multipliers.  In addition,  Figure \ref{fig:6} shows similar
ground state solutions with $M=0.5$ for different particle number $N$.

\begin{figure}[htb]
\centerline{\psfig{figure=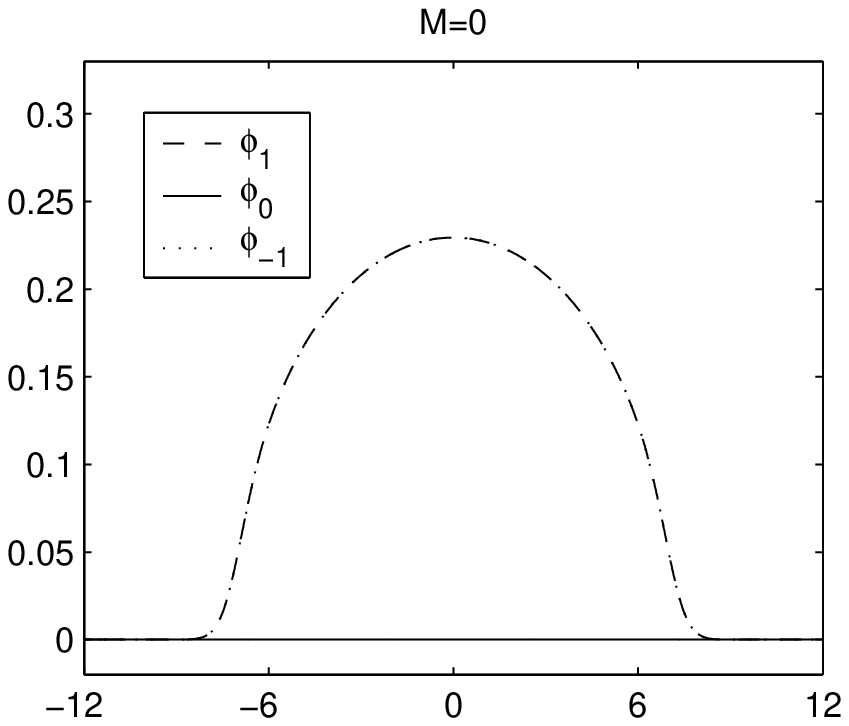,height=5cm,width=5.5cm,angle=0}
\quad \psfig{figure=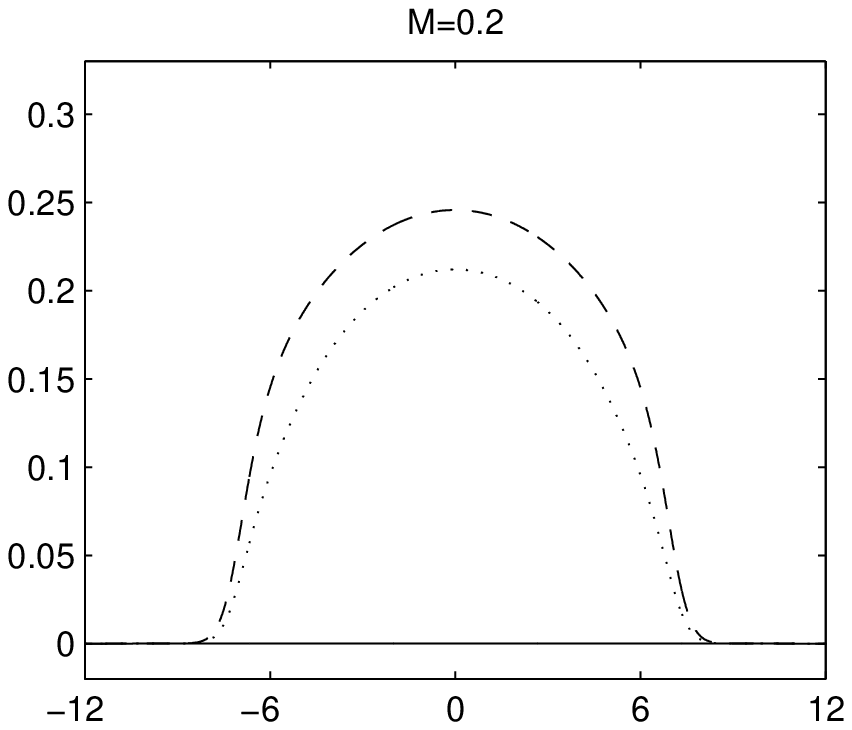,height=5cm,width=5.5cm,angle=0}}
\centerline{\psfig{figure=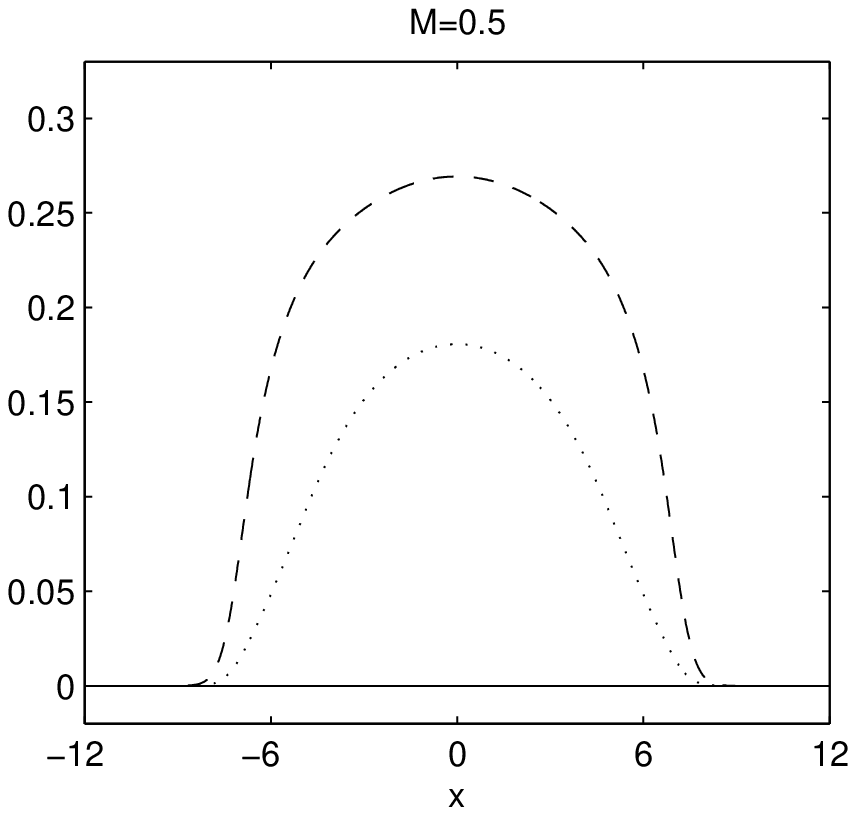,height=5.25cm,width=5.5cm,angle=0}
\quad
\psfig{figure=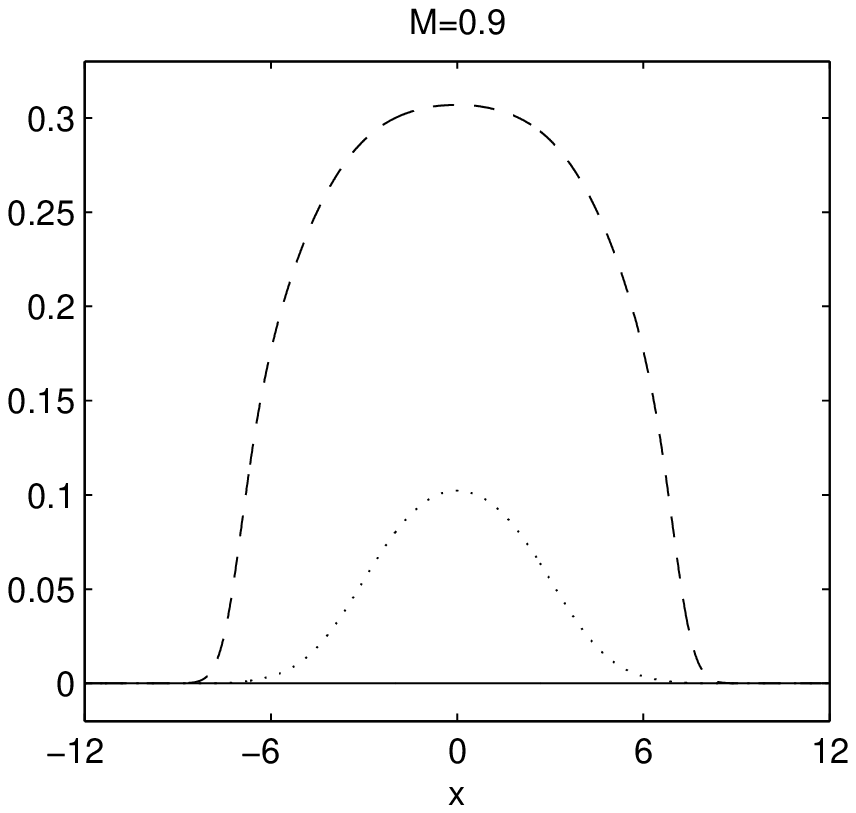,height=5.25cm,width=5.5cm,angle=0}}
\vspace{0.1cm}

\caption{Wave functions of the ground state, i.e. $\phi_1(x)$
(dashed line), $\phi_0(x)$ (solid line) and $\phi_{-1}(x)$ (dotted
line), of $^{23}$Na in Case II with fixed number of particles
$N=10^4$  for different magnetization $M=0,0.2,0.5,0.9$. }
\label{fig:5}
\end{figure}

\begin{table}[htbp]
\begin{center}
\begin{tabular}{cccc}    \hline
 $M$  & $E$  & $\mu$  & $\lambda$   \\   \hline
 0 & 15.2485 & 25.3857 & 0  \\
 0.1 & 15.2514 & 25.3847 & 0.0569  \\
 0.2 & 15.2599 & 25.3815 & 0.1142  \\
 0.3 & 15.2743 & 25.3762 & 0.1725  \\
 0.4 & 15.2945 & 25.3682 & 0.2325 \\
 0.5 & 15.3209 & 25.3572 & 0.2950  \\
 0.6 & 15.3537 & 25.3423 & 0.3611  \\
 0.7 & 15.3933 & 25.3220 & 0.4326  \\
 0.8 & 15.4405 & 25.2939 & 0.5121  \\
 0.9 & 15.4962 & 25.2527 & 0.6049 \\
 \hline
\end{tabular}
\end{center}
\caption{Ground state energy $E$ and their chemical potentials
$\mu$ and $\lambda$  for $^{23}$Na in Case II  with $N=10^4$ for
different magnetization $M$. } \label{tbl:2}
\end{table}


\begin{figure}[htb]
\centerline{\psfig{figure=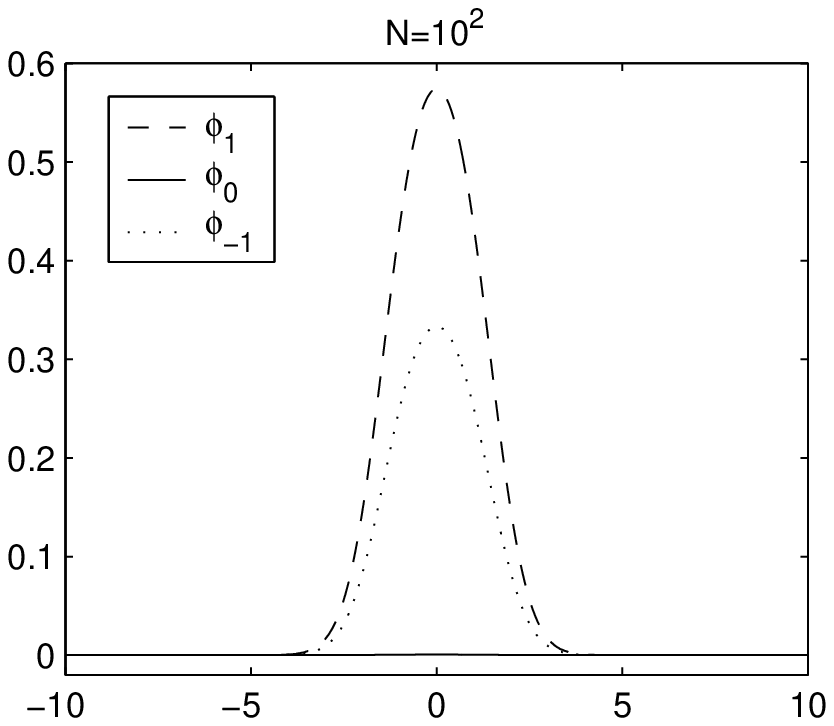,height=5cm,width=5.5cm,angle=0}
\quad \psfig{figure=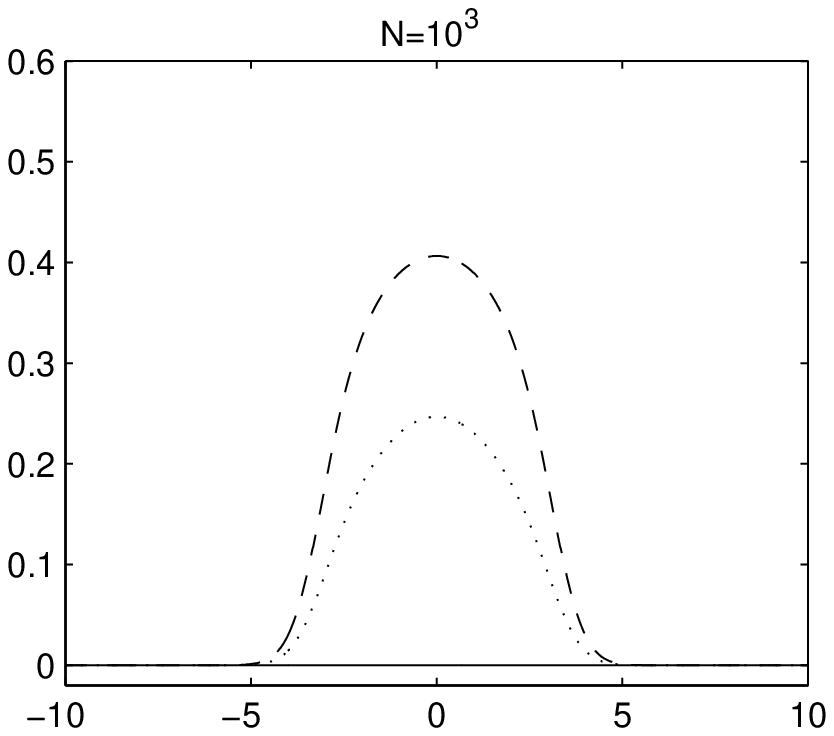,height=5cm,width=5.5cm,angle=0}}
\centerline{\psfig{figure=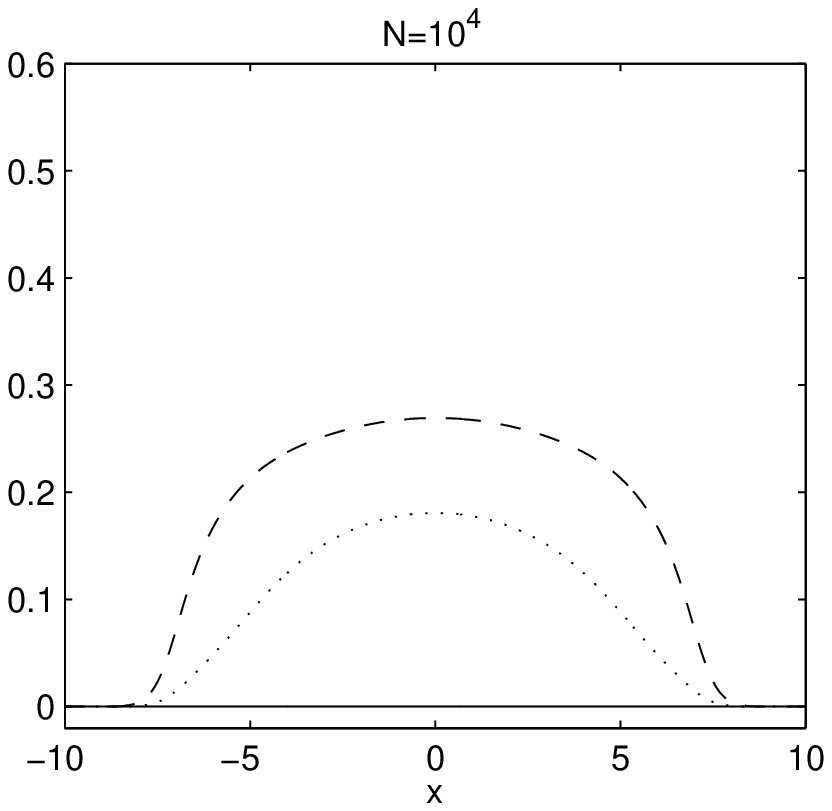,height=5.25cm,width=5.5cm,angle=0}
\quad
\psfig{figure=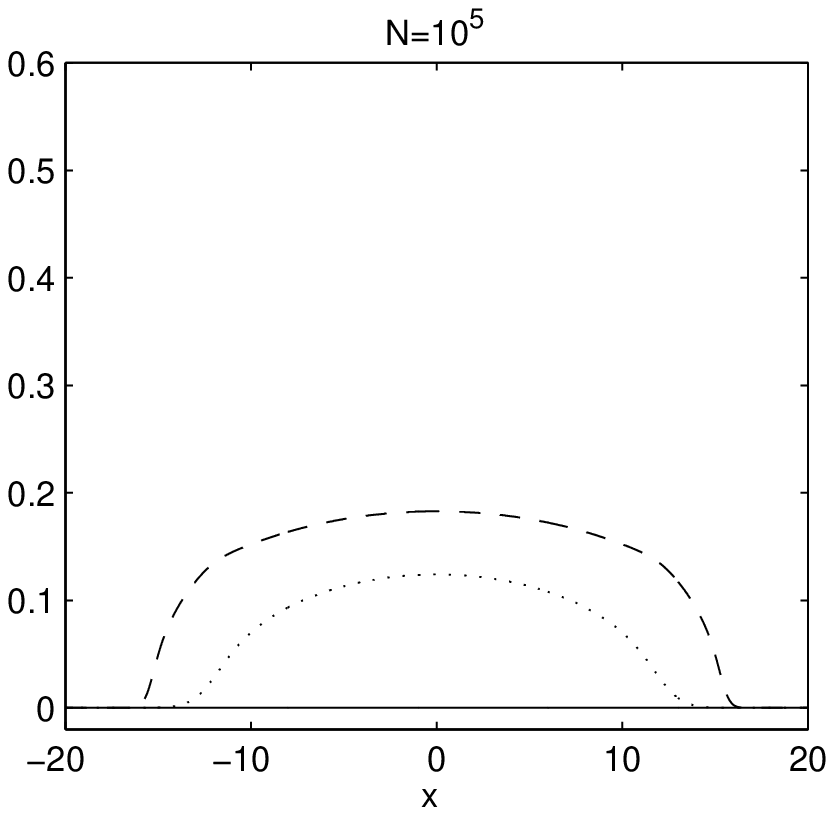,height=5.25cm,width=5.5cm,angle=0}}
\vspace{0.1cm}

\caption{Wave functions of the ground state, i.e. $\phi_1(x)$
(dashed line), $\phi_0(x)$ (solid line) and $\phi_{-1}(x)$ (dotted
line), of $^{23}$Na in Case II with magnetization $M=0.5$  for
different number of particles $N$. } \label{fig:6}
\end{figure}

Figure \ref{fig:4} plots the mass of the three
components  in the spin-1  BEC ground states with $N=10^4$ for
different magnetization $M$, and
Figure \ref{fig:3} depicts the energy and chemical potentials with $M=0.5$
for different particle number $N$.

\begin{figure}[htb]
\centerline{a)\psfig{figure=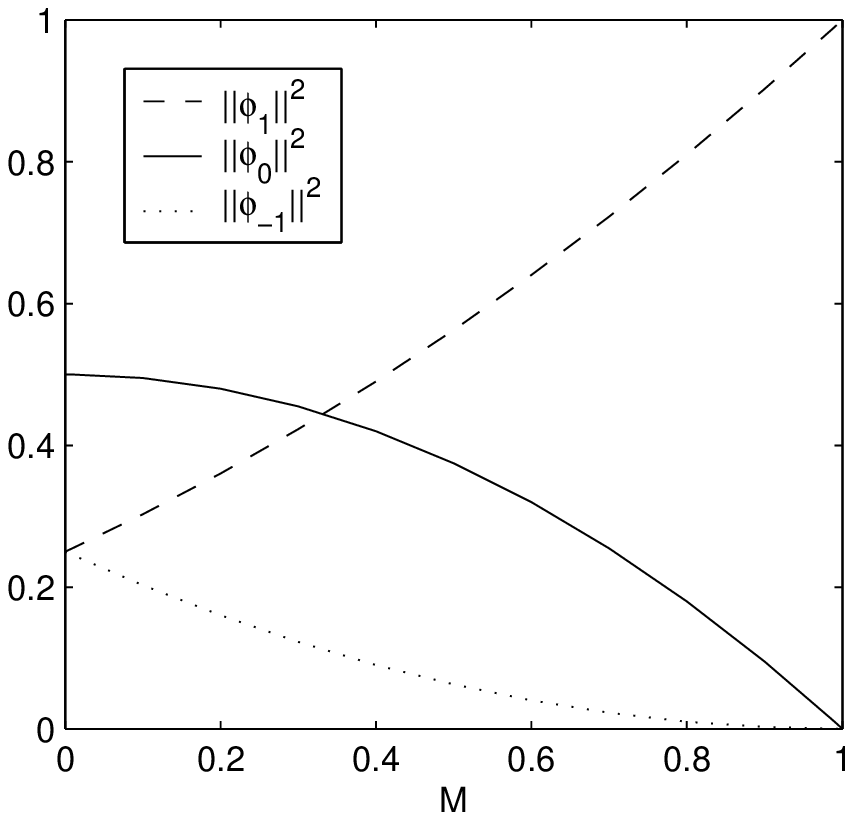,height=5.5cm,width=6cm,angle=0}
\quad
b)\psfig{figure=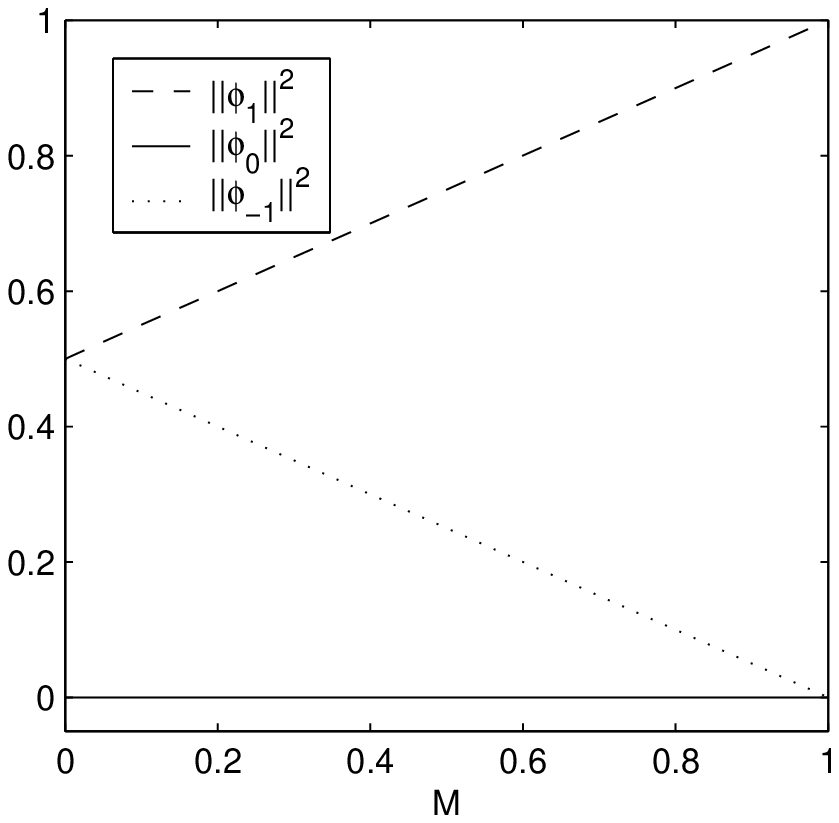,height=5.5cm,width=6cm,angle=0}}
\caption{Mass of the three components of the ground state, i.e.
$N_l=\|\phi_l\|^2$ ($l=1,0,-1$),  of spin-1 BEC with fixed number
of particles $N=10^4$ for different magnetization $0\le M< 1$.
 a) for $^{87}$Rb in case I; and b) for $^{23}$Na in case II. }
\label{fig:4}
\end{figure}

\begin{figure}[htb]
\centerline{a)\psfig{figure=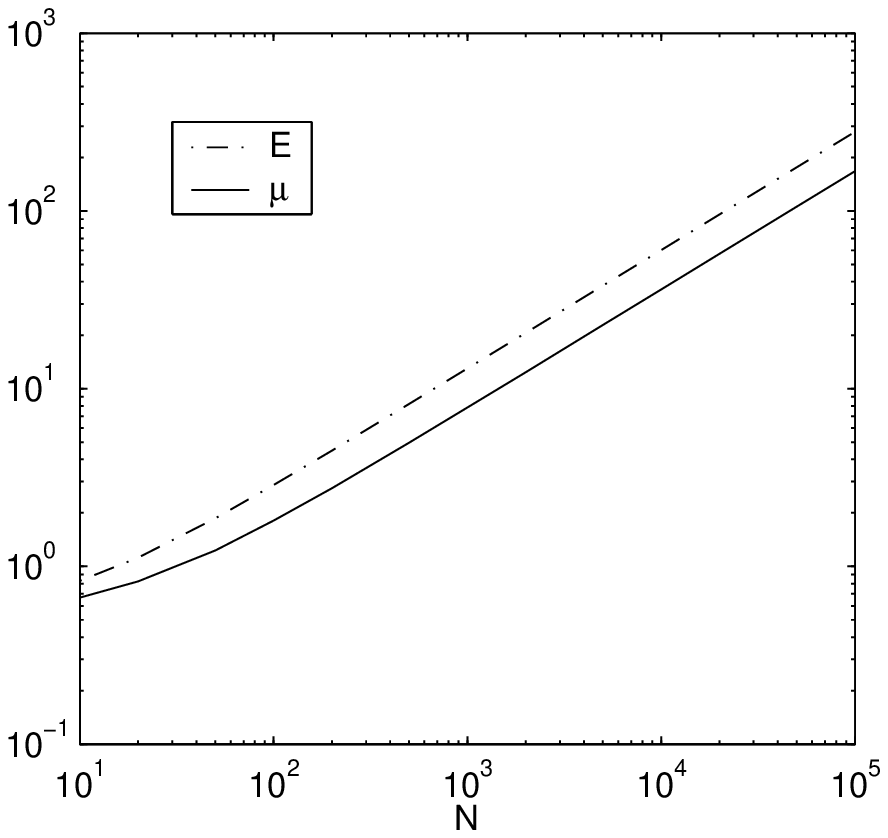,height=5.5cm,width=6cm,angle=0}
\quad
b)\psfig{figure=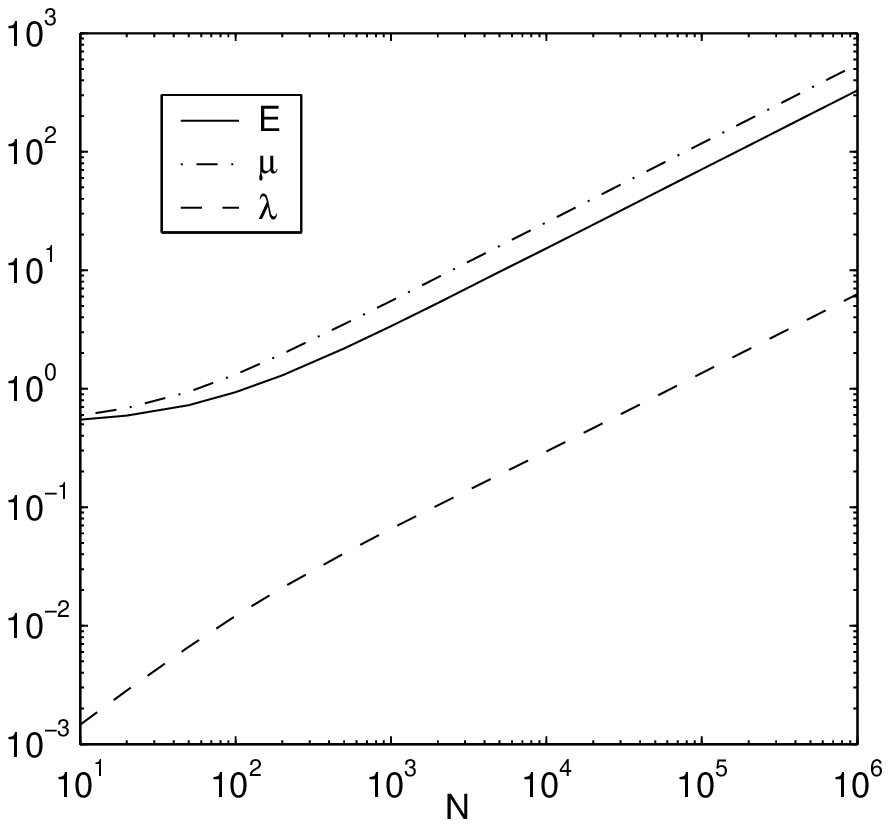,height=5.5cm,width=6cm,angle=0}
} \vspace{0.2cm}

\caption {Energy $E$ and chemical potentials $\mu$ and $\ld$ of
spin-1 BEC with fixed magnetization $M=0.5$ for different number
of particles $N$.
 a) for $^{87}$Rb in case I; and b) for $^{23}$Na in case II.
} \label{fig:3}
\end{figure}

From Figs. \ref{fig:1}-\ref{fig:5} as well as Tabs.
\ref{tbl:1}-\ref{tbl:2}, we can draw the following conclusions: (i)
For ferromagnetic interaction in the spin-1 BEC, i.e.
$\beta_s\le0$, the three components in the ground state solutions
are all positive functions (c.f. Figs. \ref{fig:1} and
\ref{fig:2}) ; while for antiferromagnetic interaction, i.e.
$\beta_s\ge0$, $\phi_1$ and $\phi_{-1}$ are positive functions and
$\phi_0\equiv 0$ (c.f. Figs. \ref{fig:5} and \ref{fig:6}). (ii)
For ferromagnetic interaction in the spin-1 BEC, i.e.
$\beta_s\le0$, for fixed number of particles $N$ in the
condensate, when the magnetization $M$ increases from $0$ to $1$,
the mass $N_1$ increases from $0.25$ to $1$, the mass $N_{-1}$
decreases from $0.25$ to $0$ and the mass $N_0$ decreases from
$0.5$ to $0$ (c.f. Fig. \ref{fig:3}a); while for antiferromagnetic
interaction, i.e. $\beta_s\ge0$, $N_1$ increases from $0.5$ to $1$,
$N_{-1}$ decreases from $0.5$ to $0$ and $N_0=0$ (c.f. Fig.
\ref{fig:3}b). (iii) For ferromagnetic interaction in the spin-1
BEC, i.e. $\beta_s\le0$, for fixed number of particles $N$ in the
condensate, the energy and chemical potentials are almost
independent of the magnetization (c.f. Tab. \ref{tbl:1}); while
for antiferromagnetic interaction, i.e. $\beta_s\ge0$, when the
magnetization $M$ increases from $0$ to $1$, the energy $E$
increases and the main chemical potential $\mu$ decreases and the
second chemical potential $\ld$ increases (c.f. Tab. \ref{tbl:2}).
In both cases, for fixed magnetization $M$, when the number of
particles $N$ increases, the energy and chemical potentials
increase (c.f. Fig. \ref{fig:4}). These observations agree with
those obtained in \cite{Bao_Wang} and \cite{You2} by different
numerical methods.

\subsection{Application in 1D with optical lattice potential}

In this subsection, our method is applied to compute the ground
state of spin-1 BEC in one dimension (1D)  with an optical lattice
potential. Again, two different interaction are considered:

\begin{itemize}

\item Case I. For $^{87}$Rb with dimensionless quantities in
(\ref{eq:dCGPE_1})-(\ref{eq:dCGPE_3}) used as: $d=1$,
$V(x)=x^2/2+25\sin^2\left(\frac{\pi x}{4}\right)$,
$\beta_n=0.0885N$
 and $\beta_s=-0.00041N$, with $N$ the total number of atoms in the condensate
and the dimensionless length unit
$a_s=2.4116\times 10^{-6}$ [m] and time unit
$t_s=0.007958$[s].

\item Case II. For $^{23}$Na with dimensionless quantities in
(\ref{eq:dCGPE_1})-(\ref{eq:dCGPE_3}) used as: $d=1$,
$V(x)=x^2/2+25\sin^2\left(\frac{\pi x}{4}\right)$,
$\beta_n=0.0241N$
 and $\beta_s=0.00075N$, with $N$ the total number of atoms in the condensate
and the dimensionless length unit $a_s=4.6896\times 10^{-6}$ [m]
and time unit $t_s=0.007958$[s].
\end{itemize}

Figure \ref{fig:1o} shows the ground state solutions of $^{87}$Rb
in Case I with $N=10^4$ for different magnetization $M$ and Table
\ref{tbl:1o} lists the corresponding ground state energies and
their Lagrange multipliers. Figure \ref{fig:2o} and Table
\ref{tbl:2o} show similar results for $^{23}$Na in Case II.

\begin{figure}[htb]
\centerline{\psfig{figure=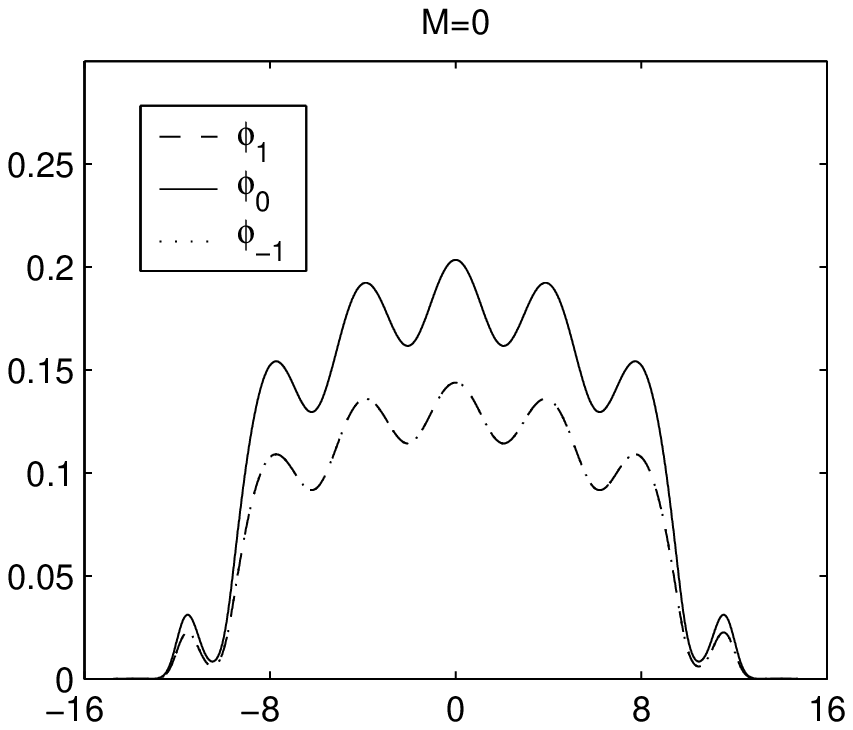,height=5cm,width=5.5cm,angle=0}
\quad
\psfig{figure=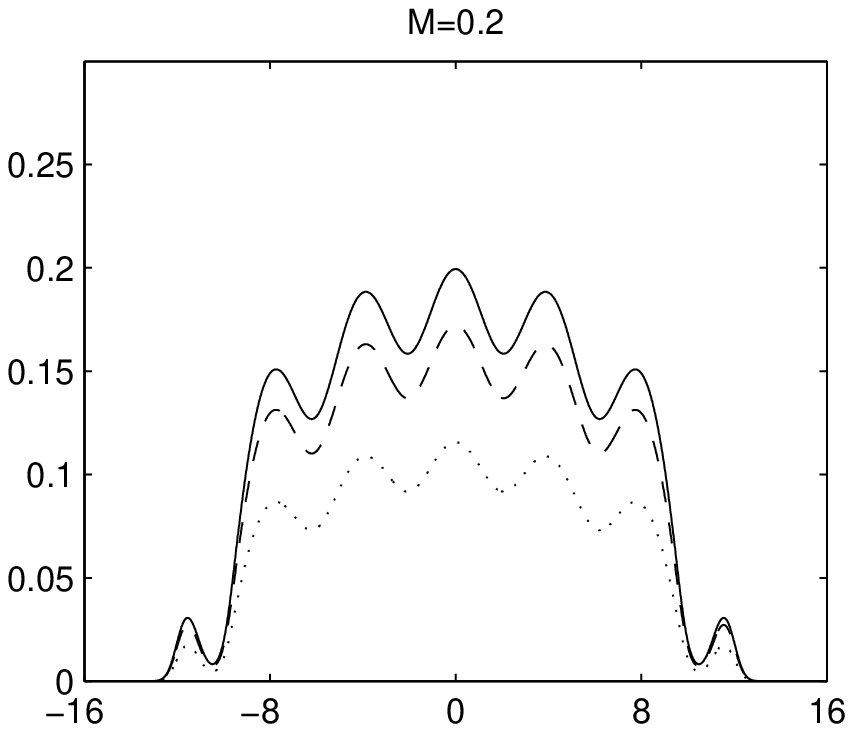,height=5cm,width=5.5cm,angle=0}}
\centerline{\psfig{figure=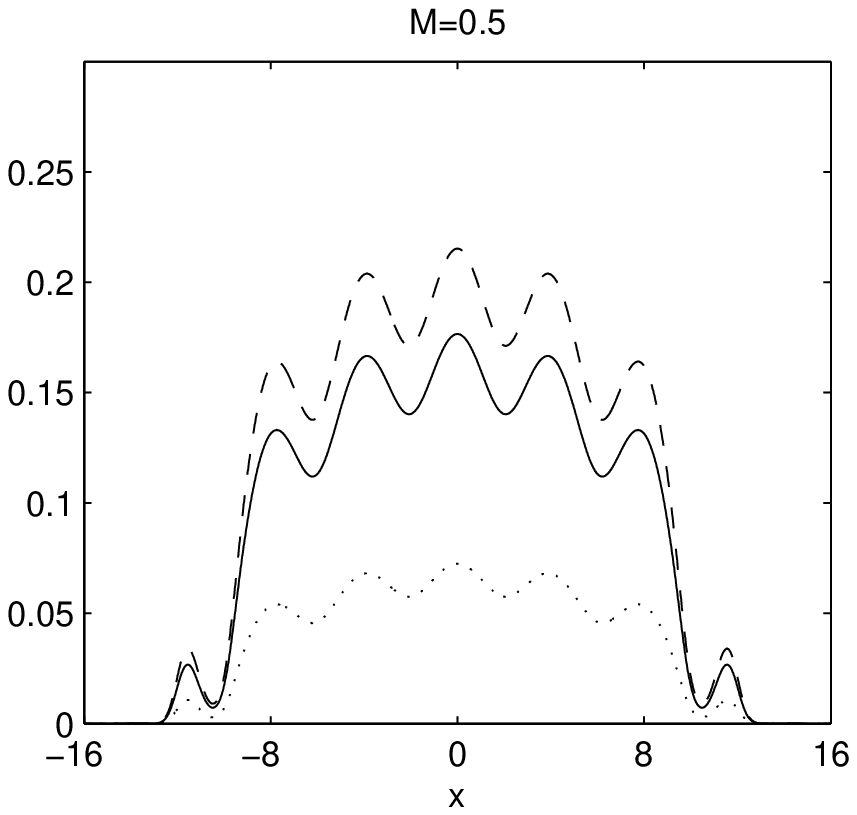,height=5.25cm,width=5.5cm,angle=0}
\quad
\psfig{figure=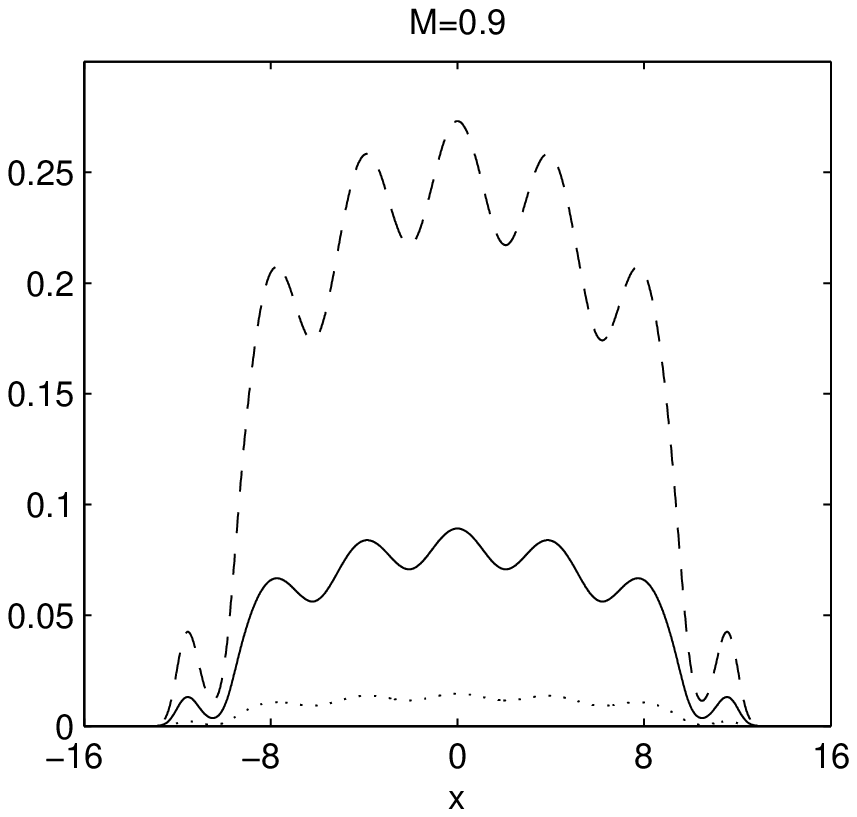,height=5.25cm,width=5.5cm,angle=0}}
\vspace{0.1cm}

\caption{Wave functions of the ground state, i.e. $\phi_1(x)$
(dashed line), $\phi_0(x)$ (solid line) and $\phi_{-1}(x)$ (dotted
line), of $^{87}$Rb in Case I with fixed number of particles
$N=10^4$  for different magnetization $M=0,0.2,0.5,0.9$ in an
optical lattice potential. } \label{fig:1o}
\end{figure}

\begin{table}[htbp]
\begin{center}
\begin{tabular}{cccc}    \hline
 $M$  & $E$  & $\mu$  & $\lambda (\times10^{-4})$  \\   \hline
 0 & 47.6944 & 73.0199 & 0  \\
 0.1 & 47.6944 & 73.0199 & 0.711   \\
 0.2 & 47.6944 & 73.0199 & 0.788   \\
 0.3 & 47.6944 & 73.0199 & 0.859  \\
 0.4 & 47.6944 & 73.0199 & 0.948   \\
 0.5 & 47.6944 & 73.0199 & 1.072   \\
 0.6 & 47.6944 & 73.0199 & 1.178   \\
 0.7 & 47.6944 & 73.0199 & 1.164   \\
 0.8 & 47.6944 & 73.0199 & 1.200   \\
 0.9 & 47.6944 & 73.0199 & 1.477  \\
 \hline
\end{tabular}
\end{center}

\caption{Ground state energy $E$ and their chemical potentials
$\mu$ and $\lambda$  for $^{87}$Rb in Case I  with $N=10^4$ for
different magnetization $M$ in an optical lattice potential.}
\label{tbl:1o}
\end{table}

\begin{figure}[htb]
\centerline{\psfig{figure=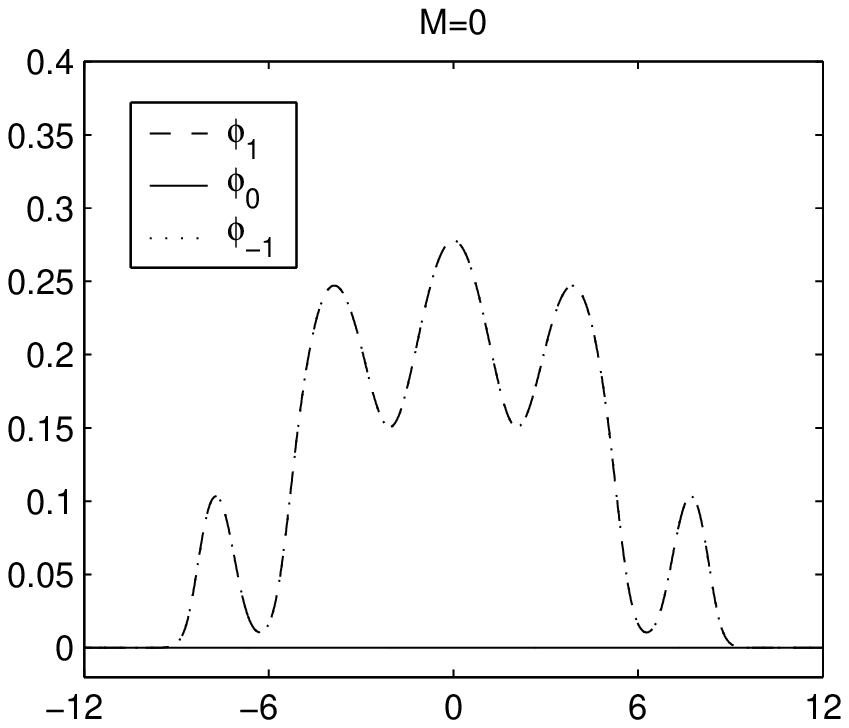,height=5cm,width=5.5cm,angle=0}
\quad
\psfig{figure=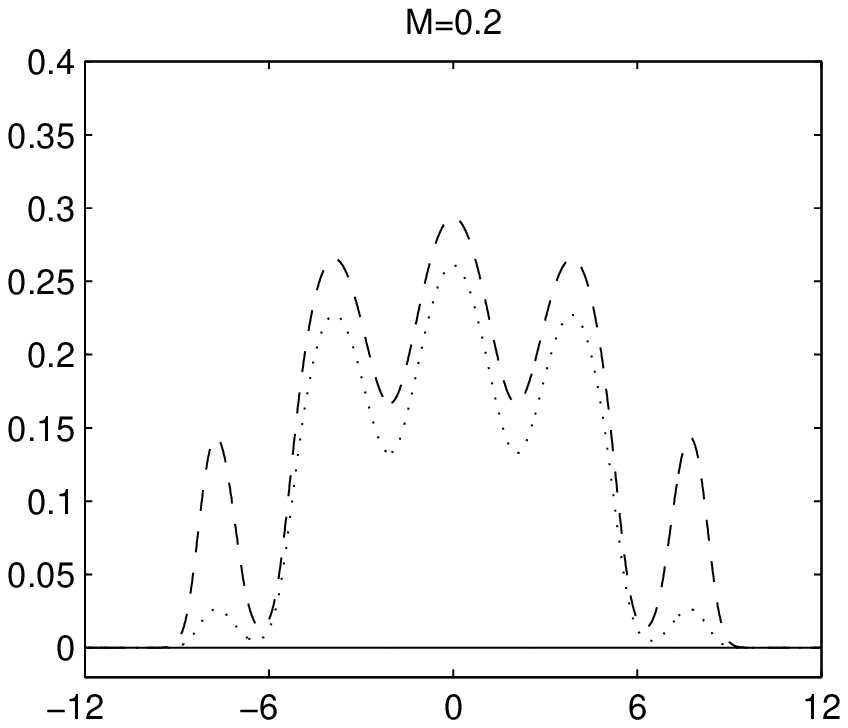,height=5cm,width=5.5cm,angle=0}}
\centerline{\psfig{figure=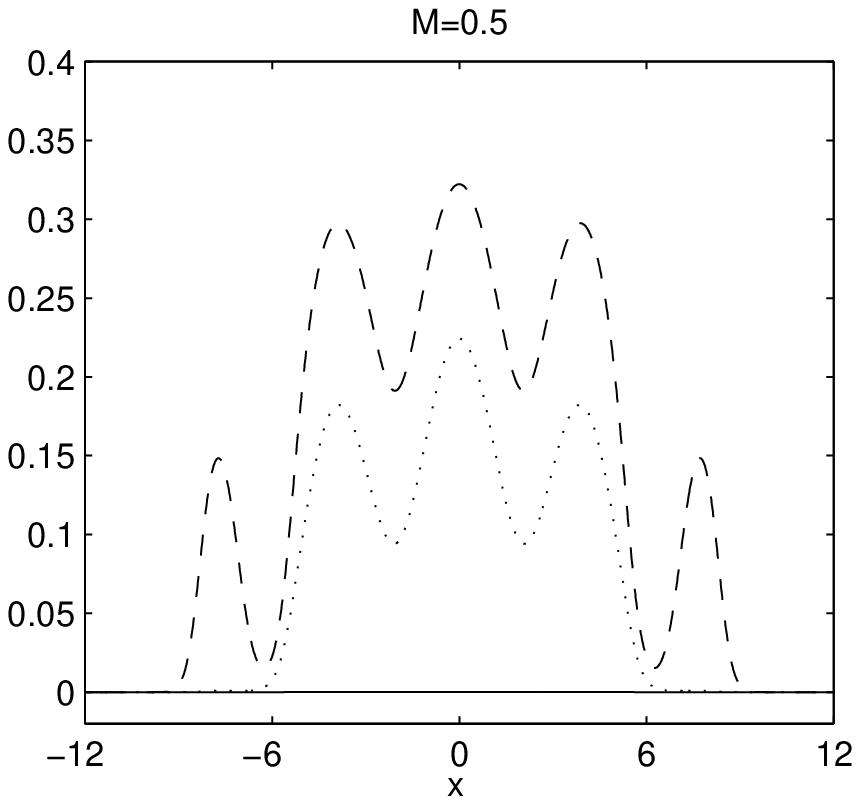,height=5.25cm,width=5.5cm,angle=0}
\quad
\psfig{figure=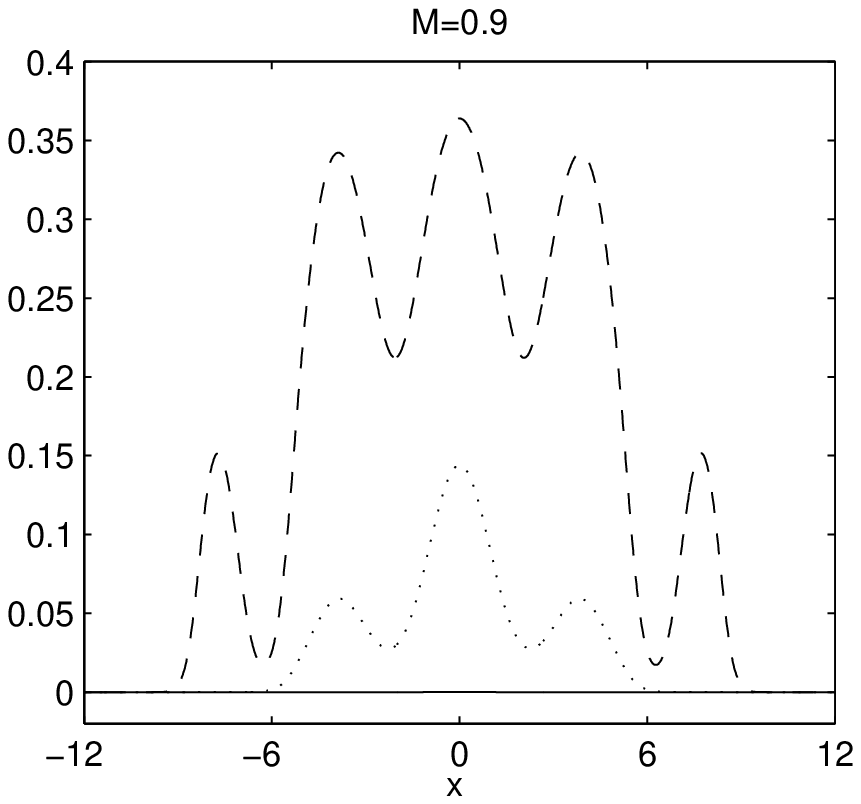,height=5.25cm,width=5.5cm,angle=0}}
\vspace{0.1cm}

\caption{Wave functions of the ground state, i.e. $\phi_1(x)$
(dashed line), $\phi_0(x)$ (solid line) and $\phi_{-1}(x)$ (dotted
line), of $^{23}$Na in Case II with $N=10^4$  for different
magnetization $M=0,0.2,0.5,0.9$ in an optical lattice potential.}
\label{fig:2o}
\end{figure}

\begin{table}[htbp]
\begin{center}
\begin{tabular}{cccc}    \hline
 $M$  & $E$  & $\mu$  & $\lambda $  \\   \hline
 0 & 25.6480 & 37.4489 & 0  \\
 0.1 & 25.6509 & 37.4476 & 0.0593  \\
 0.2 & 25.6597 & 37.4400 & 0.1197  \\
 0.3 & 25.6753 & 37.4248 & 0.1931  \\
 0.4 & 25.6983 & 37.4025 & 0.2687  \\
 0.5 & 25.7291 & 37.3775 & 0.3458  \\
 0.6 & 25.7676 & 37.3492 & 0.4252  \\
 0.7 & 25.8144 & 37.3167 & 0.5079  \\
 0.8 & 25.8696 & 37.2305 & 0.6920  \\
 0.9 & 25.9340 & 37.2305 & 0.6920  \\
 \hline
\end{tabular}
\end{center}

\caption{Ground state energy $E$ and their chemical potentials
$\mu$ and $\lambda$  for $^{23}$Na in Case II  with $N=10^4$ for
different magnetization $M$ in an optical lattice potential.}
\label{tbl:2o}
\end{table}

From Figs. \ref{fig:1o}\&\ref{fig:2o} and Tabs.
\ref{tbl:1o}\&\ref{tbl:2o}, it can be seen that our method can be used
in computing ground state of spin-1 BEC with general potential. In
addition to that, similar conclusions  as those in the
end of previous subsection can also be observed in this case.

\subsection{Applications in 3D with optical lattice potential}

In this subsection, our method is applied to compute the ground
state of spin-1 BEC in three dimensions (3D)  with an optical
lattice potential. Again, two different interaction are
considered:

\begin{itemize}

\item Case I. For $^{87}$Rb with dimensionless quantities in
(\ref{eq:dCGPE_1})-(\ref{eq:dCGPE_3}) used as: $d=3$,
$V(x)=\frac{1}{2}\left(x^2+y^2+z^2\right)
   + 100\left[\sin^2\left(\frac{\pi x}{2}\right) +
    \sin^2\left(\frac{\pi y}{2}\right) +
     \sin^2\left(\frac{\pi z}{2}\right)\right]$,
     $\beta_n=0.0880N$
 and $\beta_s=-0.00041N$, with $N$ the total number of atoms in the condensate
and the dimensionless length unit $a_s=\sqrt{\hbar/m\og_x}
=7.6262\times 10^{-7}$ [m]
and time unit $t_s=1/\og_x=7.9577\times 10^{-4}$[s] (corresponding to
physical trapping frequencies $\og_x = \og_y = \og_z = 2\pi \times 200$[Hz]).

\item Case II. For $^{23}$Na with dimensionless quantities in
(\ref{eq:dCGPE_1})-(\ref{eq:dCGPE_3}) used as: $d=3$,
$V(x)=\frac{1}{2}\left(x^2+y^2+z^2\right)
   + 100\left[\sin^2\left(\frac{\pi x}{2}\right)
   + \sin^2\left(\frac{\pi y}{2}\right) +
   \sin^2\left(\frac{\pi z}{2}\right)\right]$,
$\beta_n=0.0239N$
 and $\beta_s=0.00075N$ with $N$ the total number of atoms in the condensate
and the dimensionless length unit $a_s=1.4830 \times 10^{-6}$ [m]
and time unit $t_s=7.9577\times 10^{-4}$[s](corresponding to
physical trapping frequencies $\og_x = \og_y = \og_z = 2\pi \times 200$[Hz]).
\end{itemize}

Figure \ref{fig:3o} shows the ground state solutions  with
$N=10^4$ and $M=0.5$ for the two cases.

\begin{figure}[htb]
%
\vspace{0.1cm}

\caption{Contour plots for the wave functions of the ground state,
i.e. $\phi_1(x,y,0)$ (top row), $\phi_0(x,y,0)$ (middle row) and
$\phi_{-1}(x,y,0)$ (bottom row) with $N=10^4$ and $M=0.5$ in an
optical lattice potential. Left column: for $^{87}$Rb in Case I;
and right column: for $^{23}$Na in Case II. } \label{fig:3o}
\end{figure}

From Fig. \ref{fig:3o}, we can see that our method can be used to
compute the ground state of spin-1 BEC in 3D with general trapping
potential.

\section{Conclusions} \label{s5}

We have proposed an efficient and accurate normalized gradient
flow or imaginary time method to compute the ground state of
spin-1 Bose-Einstein condensates by introducing a third
normalization condition, in addition to the conservation of total
particle number and the conservation of total magnetization. The
condition is derived from the relationships
 between the chemical potentials of
the three spinor components together with a splitting scheme
applied to the continuous normalized gradient flows proposed to
compute the ground state of spin-1 BEC. The backward-forward
sine-pseudospectral method is applied to discretize the normalized
gradient flow  for practical computation. The ground state
solutions and fraction of each component are
 reported for both ferromagnetic and
antiferromagnetic interaction cases. The energy and chemical
potentials of the condensate are also reported.
In addition, the method may be further extended
 to other spinor condensate with higher
degree of freedom as well as spinor condensate in the
presence of external magnetic field, which will
be our future study.

  Finally, based on our extensive numerical experiments and results,
we conjecture that when $\beta_n\ge0$, $\beta_n\ge |\beta_s|$ and
$V(\bx)\ge0$ satisfying $\lim_{|\bx|\to\ift} V(\bx)\to \ift$,
there exists minimizer of the nonconvex minimization problem
(\ref{eq:minimize}). In addition, when $\beta_s<0$, positive
minimizer (the three components are positive function) is unique;
when $\beta_s>0$, nonnegative minimizer ($\phi_1$ and $\phi_{-1}$
are positive and $\phi_0\equiv 0$) is unique. Rigorous
mathematical justification are on-going.

\bigskip

\noindent{\bf Acknowledgment}. We acknowledge support from the
Ministry of Education of Singapore  grant No. R-146-000-083-112.

\bigskip
\renewcommand{\theequation}{A.\arabic{equation}}
\setcounter{equation}{0}

\begin{center}
\bf{Appendix A  Derivation of the third projection equation
(\ref{cong5}) }
\end{center}

\bigskip

In order to find the third projection or normalization equation
used in the projection step of the normalized gradient flow,
we first review the continuous normalized gradient flow (CNGF)
constructed in \cite{Bao_Wang} for computing the ground state of
spin-1 BEC in (\ref{eq:minimize}): \bea \label{cngf1}
\p_t\phi_{1}(\bx,t)&=&\left[\fl{1}{2}\nabla^2 -V(\bx)-
(\beta_n+\beta_s)\left(|\phi_1|^2+|\phi_0|^2\right)
-(\beta_n-\beta_s)|\phi_{-1}|^2\right]\phi_1\nn\\
&&-\beta_s\,\bar{\phi}_{-1}\,\phi_0^2
+\left[\mu_\Phi(t)+\ld_\Phi(t)\right]\phi_1 , \eea \bea
\label{cngf2} \p_t\phi_{0}(\bx,t)&=&\left[\fl{1}{2}\nabla^2
-V(\bx)- (\beta_n+\beta_s)\left(|\phi_1|^2+|\phi_{-1}|^2\right)
-\beta_n|\phi_{0}|^2\right]\phi_0 \nn \\
&&-2\beta_s\,\phi_{-1}\,\bar{\phi}_{0}\,\phi_1
+\mu_\Phi(t)\;\phi_0 , \eea \bea \label{cngf3}
\p_t\phi_{-1}(\bx,t)&=&\left[\fl{1}{2}\nabla^2 -V(\bx)-
(\beta_n+\beta_s)\left(|\phi_{-1}|^2+|\phi_0|^2\right)
-(\beta_n-\beta_s)|\phi_{1}|^2\right]\phi_{-1} \nn\\
&&-\beta_s\,\phi_0^2\,\bar{\phi}_{1}
+\left[\mu_\Phi(t)-\ld_\Phi(t)\right]\phi_{-1} .
\end{eqnarray}
$\mu_\Phi(t)$ and $\ld_\Phi(t)$ are chosen such that the
above CNGF is mass (or normalization) and magnetization
conservative and they are given as \cite{Bao_Wang}
 \be
\label{ldmu} \mu_\Phi(t) =\frac{R_\Phi(t) D_\Phi(t) -M_\Phi(t)
F_\Phi(t)}{N_\Phi(t)R_\Phi(t)-M_\Phi^2(t)}, \qquad
\ld_\Phi(t)=\frac{N_\Phi(t) F_\Phi(t) -M_\Phi(t)
D_\Phi(t)}{N_\Phi(t)R_\Phi(t)-M_\Phi^2(t)}, \ee with \bea
\label{NPhi} &&N_\Phi(t)=\int_{{\Bbb R}^d} \left[
|\phi_{-1}(\bx,t)|^2+
|\phi_{0}(\bx,t)|^2+|\phi_{1}(\bx,t)|^2\right]d\bx,\\
\label{MPhi} &&M_\Phi(t)=\int_{{\Bbb R}^d} \left[|\phi_1(\bx,t)|^2
-|\phi_{-1}(\bx,t)|^2\right]\;d\bx, \\
\label{RPhi} &&R_\Phi(t)=\int_{{\Bbb R}^d} \left[
|\phi_1(\bx,t)|^2 +|\phi_{-1}(\bx,t)|^2\right]\;d\bx, \eea \bea
\label{DPhi} D_\Phi(t)&=&\int_{{\Bbb R}^d}\biggl\{\sum_{l=-1}^{1}
\left(\frac{1}{2}|\nabla \phi_l|^2+V(\bx)|\phi_l|^2\right)
+2(\beta_n-\beta_s)|\phi_1|^2 |\phi_{-1}|^2 +\beta_n|\phi_0|^4\nn\\
&&+(\beta_n+\beta_s)\Bigl[|\phi_1|^4+|\phi_{-1}|^4 +2|\phi_0|^2
\left(|\phi_1|^2+|\phi_{-1}|^2\right)\Bigr]\nn\\
&&+2\beta_s\left(\bar{\phi}_{-1}\phi_0^2\bar{\phi}_{1}+
\phi_{-1}\bar{\phi}_0^2\phi_{1}\right)\biggr\}\; d\bx, \eea \bea
\label{FPhi} F_\Phi(t)&=&\int_{{\Bbb R}^d}\biggl\{\frac{1}{2}
\left(|\nabla \phi_1|^2-|\nabla \phi_{-1}|^2\right)
+V(\bx)\left(|\phi_1|^2-|\phi_{-1}|^2\right) \nn\\
&&+(\beta_n+\beta_s)\Bigl[|\phi_1|^4-|\phi_{-1}|^4
+|\phi_0|^2\left(
|\phi_1|^2-|\phi_{-1}|^2\right)\Bigr]\biggr\}\;d\bx. \eea For the
above CNGF, for any given initial data
\be\label{init}\Phi(\bx,0)=(\phi_1(\bx,0), \phi_0(\bx,0),
\phi_{-1}(\bx,0))^T :=\Phi^{(0)}(\bx), \qquad \bx\in{\Bbb R}^d,\ee
satisfying \be\label{init2}N_\Phi(t=0):=N_{\Phi^{(0)}}=1,\qquad
M_\Phi(t=0):=M_{\Phi^{(0)}}=M, \ee
 it was proven that the total mass
and magnetization are conservative and the energy is diminishing
\cite{Bao_Wang}, i.e. \[N_\Phi(t)\equiv 1, \quad
M_\Phi(t)\equiv M, \quad E\left(\Phi(\cdot, t\right) \leq
E\left(\Phi(\cdot,s)\right), \quad \hbox{for any} \ t\ge s \ge0.
\]


The normalized gradient flow
(\ref{eq:GFDN_1})-(\ref{eq:projection_3c}) can be viewed as
applying a time-splitting scheme to the CNGF
(\ref{cngf1})-(\ref{cngf3}) and the projection step
(\ref{eq:projection_1c})-(\ref{eq:projection_3c}) is equivalent to
solving the following nonlinear ordinary differential equations
(ODEs): \bea \label{ode1} &&\p_t\phi_{1}(\bx,t)=
\left[\mu_\Phi(t)+\ld_\Phi(t)\right]\phi_1 ,\\
\label{ode2} &&\p_t\phi_{0}(\bx,t)=\mu_\Phi(t)\;\phi_0,
\qquad t_{n-1}\le t\le t_{n},\qquad n\ge 1, \\
\label{ode3} &&\p_t\phi_{-1}(\bx,t)=
\left[\mu_\Phi(t)-\ld_\Phi(t)\right]\phi_{-1} .
\end{eqnarray}
The solution of the above ODEs can be expressed as
\bea\label{odes1}
&&\phi_{1}(\bx,t_{n})=\exp\left(\int_{t_{n-1}}^{t_{n}}
\left[\mu_\Phi(\tau)+\ld_\Phi(\tau)\right]\;d\tau \right)
\phi_{1}(\bx,t_{n-1}),\\
&&\phi_{0}(\bx,t_{n})=\exp\left(\int_{t_{n-1}}^{t_{n}}
\mu_\Phi(\tau)\;d\tau \right)
\phi_{0}(\bx,t_{n-1}),\\
&&\phi_{-1}(\bx,t_{n})=\exp\left(\int_{t_{n-1}}^{t_{n}}
\left[\mu_\Phi(\tau)-\ld_\Phi(\tau)\right]\;d\tau \right)
\phi_{-1}(\bx,t_{n-1}). \eea This solution suggests the following
relation between the coefficients \bea\label{coeff}
\lefteqn{\exp\left(\int_{t_{n-1}}^{t_{n}}
\left[\mu_\Phi(\tau)+\ld_\Phi(\tau)\right]\;d\tau \right)\;
\exp\left(\int_{t_{n-1}}^{t_{n}}
\left[\mu_\Phi(\tau)-\ld_\Phi(\tau)\right]\;d\tau
\right)\nn}\\[2mm]
&&=\exp\left(\int_{t_{n-1}}^{t_{n}} 2\mu_\Phi(\tau)\;d\tau
\right)=\left[\exp\left(\int_{t_{n-1}}^{t_{n}}
\mu_\Phi(\tau)\;d\tau \right)\right]^2. \eea This immediately
suggests us to propose the third normalization
equation (\ref{cong5}) to determine the projection parameters. In
fact, equation (\ref{cong5}) can be also obtained from the
relation between the chemical potentials in (\ref{chem8}) by
physical intuitions.

\bigskip
\renewcommand{\theequation}{B.\arabic{equation}}
\setcounter{equation}{0}

\begin{center}
\bf{Appendix B  Derivation of the projection parameters in
(\ref{eq:constant1})-(\ref{eq:constant2}) }
\end{center}

\bigskip

Summing (\ref{eq:constant1}) and (\ref{eq:constant2}), we get
\be\label{solp1} 2(\sg_1^n)^2 \|\phi_1(\cdot,t_n^-)\|^2
=1+M-(\sg_0^n)^2 \|\phi_0(\cdot,t_n^-)\|^2. \ee This immediately
implies \be\label{solp2} \sigma_1^{n}=\frac{\sqrt{1+M-(\sg_0^n)^2
  \|\phi_0(\cdot,t_{n}^-)\|^2}}
  {\sqrt{2}\ \|\phi_1(\cdot,t_{n}^-)\|}.
  \ee
Subtracting (\ref{eq:constant2}) from (\ref{eq:constant1}), we
obtain \be\label{solp3} 2(\sg_{-1}^n)^2
\|\phi_{-1}(\cdot,t_n^-)\|^2 =1-M-(\sg_0^n)^2
\|\phi_0(\cdot,t_n^-)\|^2. \ee Again, this immediately implies
\be\label{solp4}
  \sigma_{-1}^{n}=\frac{\sqrt{1-M-(\sg_0^n)^2
  \|\phi_0(\cdot,t_{n}^-)\|^2}}
  {\sqrt{2}\ \|\phi_{-1}(\cdot,t_{n}^-)\|}.
\ee Multiplying (\ref{solp2}) and (\ref{solp4}) and noticing
(\ref{cong5}), we get \bea\label{solp6} &&\left[1+M-(\sg_0^n)^2
  \|\phi_0(\cdot,t_{n}^-)\|^2\right]\left[1-M-(\sg_0^n)^2
  \|\phi_0(\cdot,t_{n}^-)\|^2\right]\nn \\
  &&\quad=4\|\phi_{-1}(\cdot,t_{n}^-)\|^2\;\|\phi_1(\cdot,t_{n}^-)\|^2\;
(\sg_0^n)^4. \eea Simplifying the above equation, we obtain
\bea\label{solp7}
&&\left[\|\phi_0(\cdot,t_{n}^-)\|^4-4\|\phi_{-1}(\cdot,t_{n}^-)\|^2\;
\|\phi_1(\cdot,t_{n}^-)\|^2\right](\sg_0^n)^4-2\|\phi_0(\cdot,t_{n}^-)\|^2\;
(\sg_0^n)^2\nn \\
&&\quad+(1-M^2)=0. \eea Solving the above equation and noticing
$(\sg_0^n)^2\;\|\phi_0(\cdot,t_{n}^-)\|^2\le (1-M^2)$, we get
\bea\label{solp9} (\sg_0^n)^2&=&\frac{\|\phi_0(\cdot,t_{n}^-)\|^2
-\sqrt{4(1-M^2)\|\phi_1(\cdot,t_{n}^-)\|^2
  \|\phi_{-1}(\cdot,t_{n}^-)\|^2
  + M^2\|\phi_0(\cdot,t_{n}^-)\|^4}}
  {\|\phi_0(\cdot,t_{n}^-)\|^4-4\|\phi_{-1}(\cdot,t_{n}^-)\|^2\;
\|\phi_1(\cdot,t_{n}^-)\|^2} \nn\\
&=&\frac{1-M^2}{\|\phi_0(\cdot,t_{n}^-)\|^2
+\sqrt{4(1-M^2)\|\phi_1(\cdot,t_{n}^-)\|^2
  \|\phi_{-1}(\cdot,t_{n}^-)\|^2
  + M^2\|\phi_0(\cdot,t_{n}^-)\|^4}}.
\eea Thus immediately implies the solution in
(\ref{eq:constant1}).

\bigskip
\bigskip
\renewcommand{\theequation}{C.\arabic{equation}}
\setcounter{equation}{0}

\begin{center}
\bf{Appendix C  Computing the chemical potentials $\mu$ and $\ld$}
\end{center}

After we get the ground state $\Phi$ numerically, the energy of
the ground state can be computed from the discretization of
(\ref{energy}) immediately. In order to compute the chemical
potentials numerically, different formulations can be applied.
Here we propose one of the most reliable way to compute them.
Multiplying both sides of (\ref{GPEs30}) by $\bar{\phi}_1$ and
integrate over ${\Bbb R}^d$, we get \be\label{chem21}
(\mu+\ld)\|\phi_1\|^2 =\int_{{\Bbb R}^d} \bar{\phi}_1 \; H_1
\phi_1\; d\bx:=(\phi_1,H_1\phi_1). \ee Similarly, take the same
procedure to (\ref{GPEs31}) and (\ref{GPEs32}) by multiplying
$\bar{\phi}_0$ and $\bar{\phi}_{-1}$, respectively, we obtain
\bea\label{chem22} &&\mu\|\phi_0\|^2 =\int_{{\Bbb R}^d}
\bar{\phi}_0 \; H_0 \phi_0\; d\bx:=(\phi_0,H_0\phi_0),\\
 \label{chem23} &&(\mu-\ld)\|\phi_{-1}\|^2 =\int_{{\Bbb R}^d}
\bar{\phi}_{-1} \; H_{-1} \phi_{-1}\;
d\bx:=(\phi_{-1},H_{-1}\phi_{-1}). \eea Summing (\ref{chem21}),
(\ref{chem22}) and (\ref{chem23}), noticing that the ground state
$\Phi$ satisfying the constraints (\ref{eq:S}), we get
\be\label{chem34} \mu+ M\; \ld =
(\phi_1,H_1\phi_1)+(\phi_0,H_0\phi_0)+(\phi_{-1},H_{-1}\phi_{-1}).
\ee Subtracting (\ref{chem23}) from (\ref{chem21}), we get
\be\label{chem35} M\;\mu+
\left(\|\phi_1\|^2+\|\phi_{-1}\|^2\right)\ld =(\phi_1,H_1\phi_1)
-(\phi_{-1},H_{-1}\phi_{-1}).\ee Solving the linear system
(\ref{chem34}) and (\ref{chem35}) for the chemical potentials
$\mu$ and $\ld$ as unknowns and integrating by parts to the right
hand sides,
 we have \be \label{ldmuh} \mu
=\frac{\left(\|\phi_1\|^2+\|\phi_{-1}\|^2\right) D(\Phi) -M\;
F(\Phi)}{\|\phi_1\|^2+\|\phi_{-1}\|^2-M^2}, \qquad \ld=\frac{
F(\Phi) -M\; D(\Phi)}{\|\phi_1\|^2+\|\phi_{-1}\|^2-M^2}, \ee where
\bea \label{DPhih} D(\Phi)&=&\int_{{\Bbb
R}^d}\biggl\{\sum_{l=-1}^{1} \left(\frac{1}{2}|\nabla
\phi_l|^2+V(\bx)|\phi_l|^2\right)
+2(\beta_n-\beta_s)|\phi_1|^2 |\phi_{-1}|^2 +\beta_n|\phi_0|^4\nn\\
&&+(\beta_n+\beta_s)\Bigl[|\phi_1|^4+|\phi_{-1}|^4 +2|\phi_0|^2
\left(|\phi_1|^2+|\phi_{-1}|^2\right)\Bigr]\nn\\
&&+2\beta_s\left(\bar{\phi}_{-1}\phi_0^2\bar{\phi}_{1}+
\phi_{-1}\bar{\phi}_0^2\phi_{1}\right)\biggr\}\; d\bx, \eea \bea
\label{FPhih} F(\Phi)&=&\int_{{\Bbb R}^d}\biggl\{\frac{1}{2}
\left(|\nabla \phi_1|^2-|\nabla \phi_{-1}|^2\right)
+V(\bx)\left(|\phi_1|^2-|\phi_{-1}|^2\right) \nn\\
&&+(\beta_n+\beta_s)\Bigl[|\phi_1|^4-|\phi_{-1}|^4
+|\phi_0|^2\left(
|\phi_1|^2-|\phi_{-1}|^2\right)\Bigr]\biggr\}\;d\bx. \eea Thus the
chemical potentials $\mu$ and $\ld$ can be computed numerically
from the discretization of (\ref{ldmuh}), (\ref{DPhih}) and
(\ref{FPhih}).


\end{document}